# Trans-Bifurcation Prediction of Dynamics in terms of Extreme Learning Machines with Control Inputs

Satoru Tadokoro[1], Akihiro Yamaguchi[2], Takao Namiki[1, 3] and Ichiro Tsuda[4]
[1]Education and Research Center for Mathematical and Data Science, Hokkaido University, Sapporo, Japan
[2]Department of Information and Systems Engineering, Fukuoka Institute of Technology, Fukuoka, Japan
[3]Department of Mathematics. Faculty of Science, Hokkaido University, Sapporo, Japan
[4]AIT Center, Sapporo City University, Sapporo, Japan

**Abstract**: By extending the extreme learning machine by additional control inputs, we achieved almost complete reproduction of bifurcation structures of dynamical systems. The learning ability of the proposed neural network system is striking in that the entire structure of the bifurcations of a target one-parameter family of dynamical systems can be nearly reproduced by training on transient dynamics using only a few parameter values. Moreover, we propose a mechanism to explain this remarkable learning ability and discuss the relationship between the present results and similar results obtained by Kim et al.

**Predicting the behaviors of dynamical systems is critical in various scientific fields. The recent advancements in machine learning have enabled the prediction of dynamical system trajectories for fixed parameter values. However, the accuracy of predicting behavioral changes due to changes in parameter values remains insufficient. In this study, we extend the extreme learning machine, a neural network-based technique, to enhance its capabilities for predicting behavioral changes in response to changes in parameter values. We subsequently evaluate the predictive performance of this extended model for a variety of dynamical systems, thereby elucidating its underlying mechanisms.**

## I. INTRODUCTION

Differential or difference equations serve as foundational models for capturing diverse dynamic behaviors in nonlinear dynamical systems. Understanding their bifurcation structures is crucial for elucidating the mechanisms underlying various complex dynamics. Recent advancements in machine learning, specifically in the domain of reservoir computing [1, 2], have facilitated the accurate reproduction of short-term trajectories and ergodic properties of dynamical systems [3, 4]. However, for a comprehensive understanding of an entire target system through learning, the model should have the ability to not only replicate the observed behaviors of a target system but also predict the global bifurcation structure. Here, the reservoir computer consists of the following layers: input, internal (i.e. hidden), and output. Typically, a hidden layer possesses random neural networks, and synaptic learning is performed only at the output layer with the help of a supervisor. Notably, echo state networks have echo state properties that maintain input information for at least some time.

Recently, attempts have been made to predict bifurcations using neural networks trained on the time-series data of dynamical systems [5] [6] [7] [8] [9] [10]. Notably, Kim et al. [11] introduced a pioneering extension of an echo state network by incorporating additional neuron types into the input layer. These neurons receive control inputs that facilitate post-learning modulation of the bifurcation parameters, which is also the subject of learning. Hereafter, we adopt Kim's model for echo state network with control inputs (ESNC). They trained the ESNC using trajectories of the target dynamical system with only a few values of the bifurcation parameter, thereby demonstrating the capabilities of ESNC, such as replicating the system behaviors, even at parameter values significantly different from those used during training. For instance, they demonstrated that ESNC, which was trained only on a small dataset of dynamical trajectories in the Lorenz system, converges to some fixed points, exhibited chaotic trajectories and even bifurcation structures at unlearned bifurcation parameter values when only modulating the control input. Here, such predictions as those of Kim et al. are referred to as *trans-bifurcation predictions.*

The extreme learning machine (ELM) [12] is a feedforward neural network known for its low computational cost. Specifically, ELM is constructed using three-layer neural networks, with optimization performed only on the weight matrix of the output layer. Exploring whether ELM can achieve trans-bifurcation prediction when extended in a manner similar to that of Kim et al [11] is of interest.

In the present study, we extended the input layer of ELM by adding supplementary neurons that receive control inputs in a manner similar to that of Kim et al [11] . This allowed us to propose an extended mechanism for trans-bifurcation prediction in extreme learning machine with control inputs (ELMC), thereby showing that this extended model has trans-bifurcation predictive capability, similar to ESNC.

## II. EXTREME LEARNING MACHINE WITH CONTROL INPUTS

Fig. 1 illustrates the ELMC architecture. The input layer consists of two distinct types of inputs: $\boldsymbol{x} \in \mathbb{R}^{N_x}$ and $\boldsymbol{c} \in \mathbb{R}^{N_c}$, where $N_x$ and $N_c$ represent the number of dynamical and control inputs, respectively. In ELMC, in the process of learning, the dynamics of a target dynamical system, that is, the time-series data of the system's state variables, are fed into input $\boldsymbol{x}$, whereas the control signals corresponding to the bifurcation parameter values are fed into $\boldsymbol{c}$. The dimension of the control input

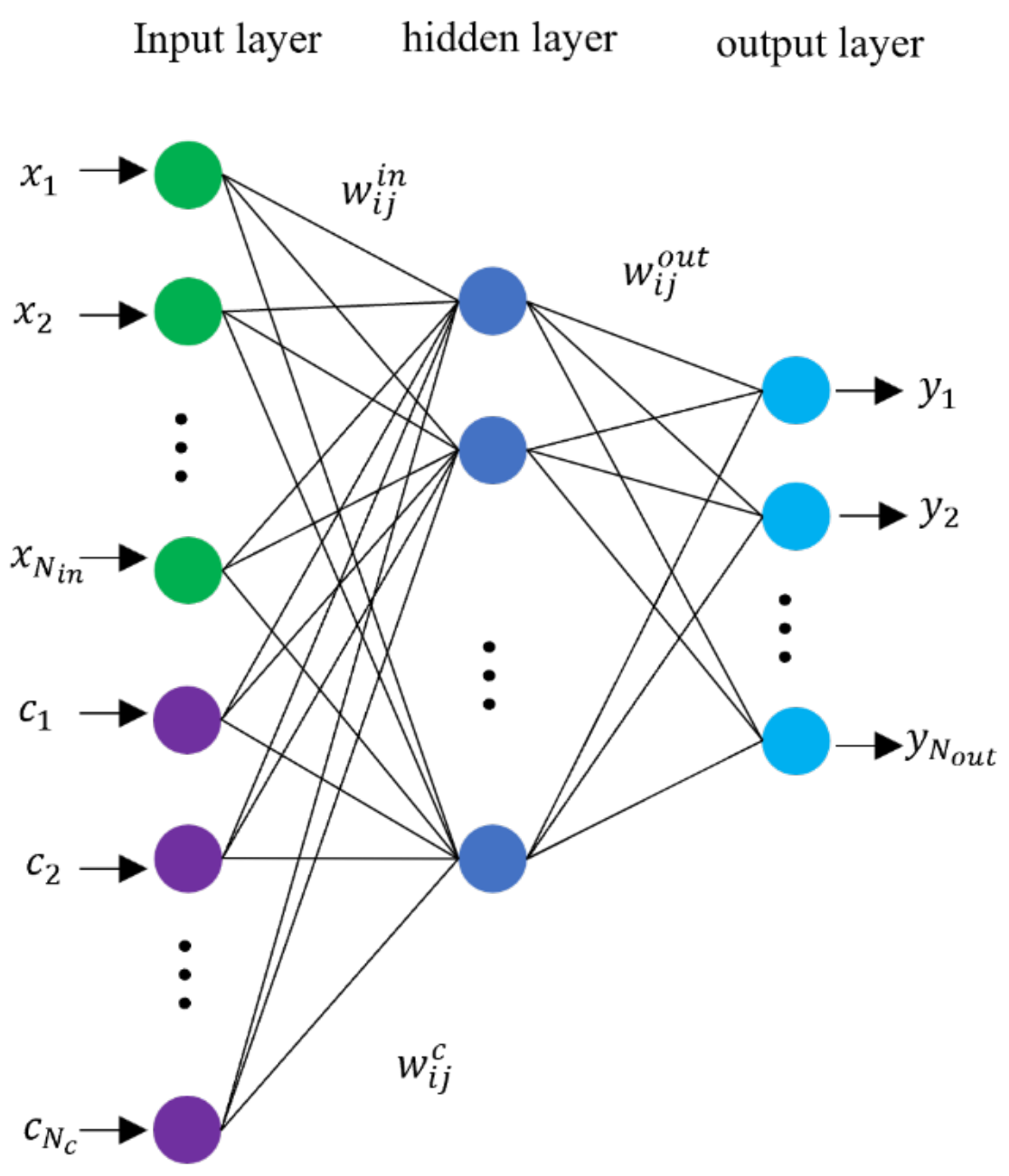

Fig. 1 Architecture of extreme learning machine with control inputs

$N_c$ was set equal to the number of bifurcation parameters of interest.

The states of the $N_h$ hidden layer neurons $\boldsymbol{h} \in \mathbb{R}^{N_h}$ are expressed as follows:

$$\boldsymbol{h} = \tanh(\alpha \boldsymbol{W}^{in}\boldsymbol{u} + \beta \boldsymbol{W}^{c}\boldsymbol{c} + \boldsymbol{\theta}) \quad (1)$$

where $\boldsymbol{W}^{in} \in \mathbb{R}^{N_h \times N_x}$ and $\boldsymbol{W}^{c} \in \mathbb{R}^{N_h \times N_c}$ denote the weight matrices connecting $\boldsymbol{x}$ and $\boldsymbol{c}$ to the hidden layer neurons, respectively. The scale factors $\alpha$ and $\beta$ regulate the influence of these weight matrices. The vector $\boldsymbol{\theta} \in \mathbb{R}^{N_h}$ represents a bias. The elements of $\boldsymbol{W}^{in}$ and $\boldsymbol{W}^{c}$ are randomly chosen from a uniform distribution ranging between -1 and 1. In accordance with Kim et al. [11] , the elements of $\boldsymbol{\theta}$ are determined such that the states $\boldsymbol{h}$ are distributed within the interval [-1, -0.8] or [0.8, 1] when both the dynamical and control inputs are zero in Eq. (1). By setting the bias term in this manner, the majority of activation variable values $(\alpha \boldsymbol{W}^{in}\boldsymbol{u} + \beta \boldsymbol{W}^{c}\boldsymbol{c} + \boldsymbol{\theta})$ are maintained within the nonlinear region of the tanh function. The outputs $\boldsymbol{y} \in \mathbb{R}^{N_{out}}$ generated by ELMC are calculated as a linear combination of the states of hidden layer neurons, defined as:

$$\boldsymbol{v} = \boldsymbol{W}^{out}\boldsymbol{h}, \tag{2}$$

where $N_{out}$ and $\boldsymbol{W}^{out} \in \mathbb{R}^{N_{out} \times N_h}$ represent the number of outputs and the output weight matrix, respectively. The matrix is determined to minimize the difference between output $\boldsymbol{v}$ and target output $\boldsymbol{d}$. Specifically, a dynamical dataset $\boldsymbol{U}$, a control dataset $\boldsymbol{C}$, and target dataset $\boldsymbol{D}$, each comprising $N$ distinct data points, are provided.

$$\begin{aligned} \boldsymbol{U} &= \big(\boldsymbol{u}(1), \cdots, \boldsymbol{u}(N)\big) \\ \boldsymbol{C} &= \big(\boldsymbol{c}(1), \cdots, \boldsymbol{c}(N)\big) \ , \\ \boldsymbol{D} &= \big(\boldsymbol{d}(1), \cdots, \boldsymbol{d}(N)\big) \end{aligned} \tag{3}$$

By substituting columns $\boldsymbol{u}(n)$ of $\boldsymbol{U}$ and $\boldsymbol{c}(n)$ of $\boldsymbol{C}$ for $\boldsymbol{x}$ and $\boldsymbol{c}$, respectively, in Eq. (1), the matrix $\boldsymbol{H}$ of the hidden layer neuron states is constructed as follows:

$$\begin{aligned} \boldsymbol{H} &= \big(\boldsymbol{h}(1), \cdots, \boldsymbol{h}(N)\big) \\ \boldsymbol{h}(n) &= \tanh\big(\alpha \boldsymbol{W}^{in}\boldsymbol{u}(n) + \beta \boldsymbol{W}^{c}\boldsymbol{c}(n) + \boldsymbol{\theta}\big) \end{aligned} \tag{4}$$

Then, matrix $\boldsymbol{W}^{out}$ is determined by minimizing the matrix 2-norm $||\boldsymbol{D} - \boldsymbol{W}^{out}\boldsymbol{H}||^2$, expressed as

$$\boldsymbol{W}^{out} = \arg\min_{\boldsymbol{W}} ||\boldsymbol{D} - \boldsymbol{W}\boldsymbol{H}||^2 \tag{5}$$

This minimization was accomplished using the Moore-Penrose pseudo-inverse matrix.

## III. TRAINING DATASETS OF A TARGET DYNAMICAL SYSTEM

For clarity, we explain the procedure for constructing the training dataset to examine the dynamical behaviors of one-parameter families of discrete dynamical systems and their bifurcation structures. The target dynamical system is expressed as follows:

$$\boldsymbol{u}(n+1) = \boldsymbol{f}\big(a, \boldsymbol{u}(n)\big), \tag{6}$$

where $a$ is the bifurcation parameter. As only one parameter was varied in this study, $N_c = 1$.

A time series of $\boldsymbol{u}$ was generated starting from $N_{init}$ different initial conditions for each of the two values of the bifurcation parameter $a = a_1$ and $a = a_2$. Denoting the state vector at time $n$ as $a = a_i$ and the $j$-th initial condition as $\boldsymbol{u}_i^j(n)$, the time series is represented by

$$\left(\boldsymbol{u}_i^j(1), \boldsymbol{u}_i^j(2), \cdots, \boldsymbol{u}_i^j(n), \cdots\right) \tag{7}$$

The input data matrix $\boldsymbol{U}$ is constructed by horizontally concatenating $N_{init}$ time series data, each with length $N$ of the bifurcation parameter values, $a = a_1$ and $a = a_{2,}$, in the following order:

$$\begin{aligned}\boldsymbol{X} = (&\boldsymbol{u}_1^1(1), \boldsymbol{u}_1^1(2), \cdots, \boldsymbol{u}_1^1(N), \boldsymbol{u}_1^2(\mathbf{1}), \boldsymbol{x}_1^2(2), \cdots, \boldsymbol{u}_1^2(N), \cdots, \boldsymbol{u}_1^{N_{init}}(1), \boldsymbol{u}_1^{N_{init}}(2), \cdots, \boldsymbol{u}_1^{N_{init}}(N),\\ &\boldsymbol{u}_2^1(1), \boldsymbol{u}_2^1(2), \cdots, \boldsymbol{u}_2^1(N), \boldsymbol{u}_2^2(1), \boldsymbol{u}_2^2(2), \cdots, \boldsymbol{u}_2^2(N), \cdots, \boldsymbol{u}_2^{N_{init}}(1), \boldsymbol{u}_2^{N_{init}}(2), \cdots, \boldsymbol{u}_2^{N_{init}}(N)\Big)\end{aligned} \tag{8}$$

The aim was to predict state vector $\boldsymbol{u}$ at time $n+1$ based on $\boldsymbol{u}$ at time $n$. The resulting target data matrix $\boldsymbol{D}$ corresponding to $\boldsymbol{U}$ is expressed as follows:

$$\begin{aligned}\boldsymbol{D} = (&\boldsymbol{u}_1^1(2), \boldsymbol{u}_1^1(3), \cdots, \boldsymbol{u}_1^1(N+1), \boldsymbol{u}_1^2(2), \boldsymbol{u}_1^2(3), \cdots, \boldsymbol{u}_1^2(N+1), \cdots, \boldsymbol{u}_1^{N_{init}}(2), \boldsymbol{u}_1^{N_{init}}(3), \cdots, \boldsymbol{u}_1^{N_{init}}(N+1),\\ &\boldsymbol{u}_2^1(2), \boldsymbol{u}_2^1(3), \cdots, \boldsymbol{u}_2^1(N+1), \boldsymbol{u}_2^2(2), \boldsymbol{u}_2^2(3), \cdots, \boldsymbol{u}_2^2(N+1), \cdots, \boldsymbol{u}_2^{N_{init}}(2), \boldsymbol{u}_2^{N_{init}}(3), \cdots, \boldsymbol{u}_2^{N_{init}}(N+1)\Big),\end{aligned} \tag{9}$$

Data matrices $\boldsymbol{U}$ and $\boldsymbol{D}$ have column length $2 \times N_{init} \times N$. Notably, the columns in the first half of $\boldsymbol{U}$ and $\boldsymbol{D}$ correspond to the bifurcation parameter $a = a_1$, whereas the columns in the second half correspond to $a = a_2$.

The control signal data matrix $\boldsymbol{C}$ is formulated as follows. All $N_{init} \times N$ elements in the first and second halves of $\boldsymbol{C}$ are set to constants $c_1$, and $c_2$, respectively. There are no specific constraints on the values of $c_1$ and $c_2$.

Determining $\boldsymbol{W}^{out}$ using dataset $(\boldsymbol{U}, \boldsymbol{C}, \boldsymbol{D})$, an autonomous dynamical system can be created by closing the feedback loop and substituting the output $\boldsymbol{x}(n) = \boldsymbol{W}^{out}\boldsymbol{h}(n-1)$ for the input:

$$\boldsymbol{u}(n+1) = \boldsymbol{W}^{\boldsymbol{out}} \tanh\left(\alpha \boldsymbol{W}^{\boldsymbol{in}} \boldsymbol{u}(n) + \beta \boldsymbol{W}^{\boldsymbol{c}} \boldsymbol{c} + \boldsymbol{\theta}\right) \tag{10}$$

Here, the temporal index $n$ of the control signal is omitted because we adopted a constant control input in this autonomous system. Let us define $\boldsymbol{f}^{ELMC}(c, \boldsymbol{u})$ on the right side of Eq. (10) as:

$$\boldsymbol{f}^{ELMC}(c, \boldsymbol{u}) = \boldsymbol{W}^{\boldsymbol{out}} \tanh\left(\alpha \boldsymbol{W}^{\boldsymbol{in}} \boldsymbol{u} + \beta \boldsymbol{W}^{\boldsymbol{c}} \boldsymbol{c} + \boldsymbol{\theta}\right), \tag{11}$$

and then

$$\boldsymbol{u}(n+1) = \boldsymbol{f}^{ELMC}\left(c, \boldsymbol{u}(n)\right) \tag{12}$$

This is a discrete dynamical system. By providing the relationship between the control parameter $c$ and bifurcation parameter $a$, as in Eq. (13), the dynamical system approximates the target dynamical system.

$$a(c) = a_1 + \frac{(a_2 - a_1)}{(c_2 - c_1)}(c - c_1) \tag{13}$$

The detailed derivation of this formula is provided in Section IV.

## IV. REALIZATION OF ELMC IN DISCRETE DYNAMICAL SYSTEMS

### A. Case of logistic map

In this study, for the first time, we realize ELMC with the logistic map $x(n+1) = ax(n)(1 - x(n))$. The training dataset was constructed using the method described in the previous section, in which a typical case of $a_1 = 3.3$, $a_2 = 3.4$, $c_1 = 0$, $c_2 = 1$, $N_{init} = 50$ and $N = 5$ was used. $N_{init}$ distinct initial values of $x(0)$ were sampled uniformly over an open interval (0, 1). Each trajectory in this training dataset exhibits transient behavior, ultimately converging to a period-two cycles. For the ELMC parameters, we used $N_h = 20$, $\alpha = 2.0$, and $\beta = 0.00002$. These parameter values although special are also typical of the learning capability of the system. Indeed, we confirmed that the approximation of dynamical systems holds for a wide range of parameter values.

Fig. 2 illustrates the time series $\{x(n)\}$ of the logistic map, along with the predicted time series generated by ELMC for three different parameter values: $a = 3.5$ ($c$ =2.0), $a = 3.62$ ($c$ =3.2), and $a = 3.8276$ ($c = 5.2756$). The top panel shows that ELMC accurately predicts the periodic trajectory for $a = 3.5$. The middle and bottom panels show that ELMC provides reasonable short-term predictions even in the chaotic regime ($a = 3.62$ and $3.8276$, respectively). Although ELMC is unable to achieve long-term predictions of chaotic trajectories, it effectively reproduces certain chaotic behavior characteristics, as illustrated in Fig. 3. Panels (a) and (d) represent scatter plots of the time-series data at the same parameter values as those in the middle and bottom panels of Fig. 2.

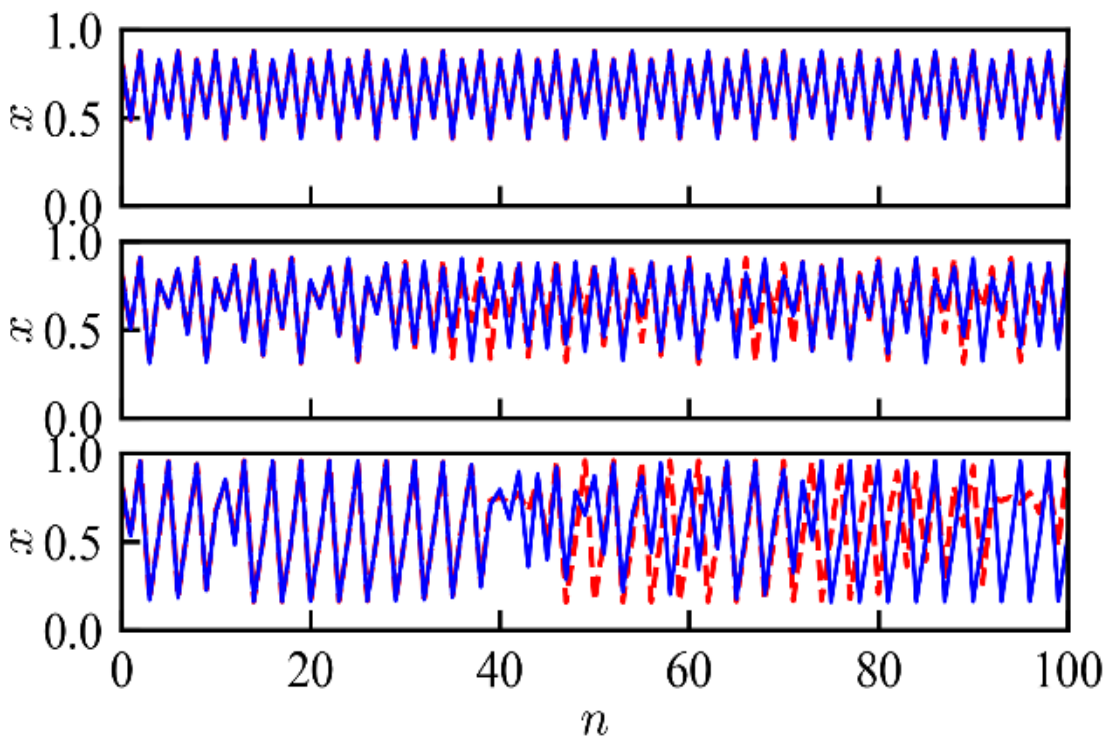

Fig. 2 Comparison of predicted (blue line) and actual (red line) time series of $x(n)$ of the logistic map for three different parameter values: $a = 3.5$ ($c$ =2.0) in the top panel; $a = 3.62$ ($c$ =3.2) in the middle panel; and $a = 3.8276$ ($c = 5.2756$) in the bottom panel.

These figures indicate that ELMC closely replicates the characteristic patterns in the time series, each displaying an association with band chaos in Panel (a) and intermittency in Panel (d). Furthermore, the invariant measures of these chaotic trajectories were calculated for both the logistic map and ELMC, as shown in Panels (b), (c), (e), and (f). Comparisons of the invariant measures of ELMC with those of the original logistic map in cases of band chaos and intermittency demonstrated that ELMC can even reproduce the ergodic property of the system, which implies that long-term dynamic behaviors can be reproduced as an attractor.

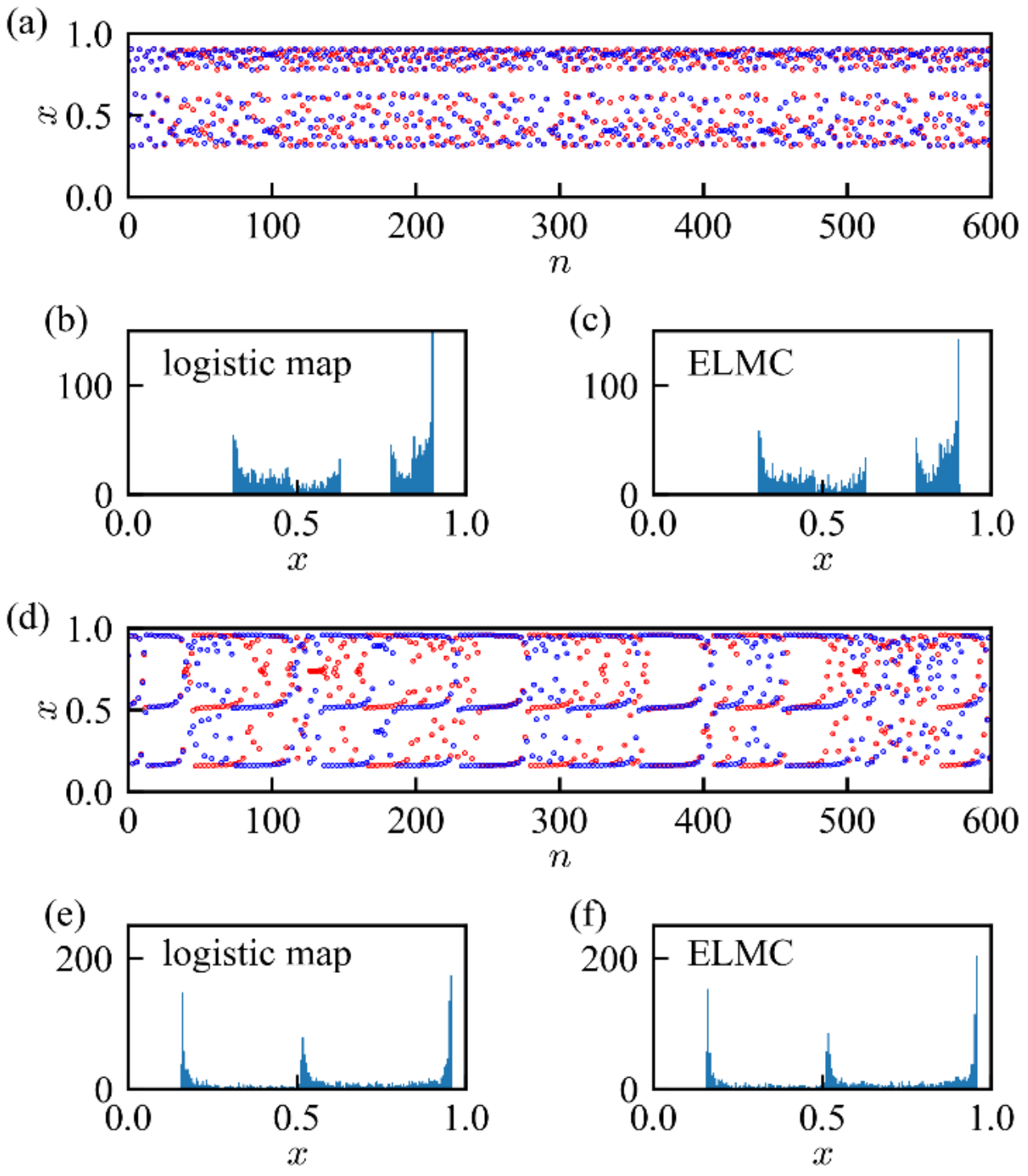

Fig. 3 Time series of x(n) of the logistic map (depicted by red dots) overlaid on the predicted time series generated by ELMC (shown by blue dots) in (a) and (d). The corresponding invariant measures are shown in (b), (c), (e) and (f). Panels (a) , (b) and (c) are for the case of $a = 3.62$ ($c = 2.7$), and Panels (d), (e) and (f) for $a = 3.8276$ ($c = 5.276$), respectively.

Fig. 4 compares the bifurcation diagram derived using ELMC for the logistic map with the original bifurcation diagram. As depicted in the figure, the ELMC successfully generates an entire bifurcation

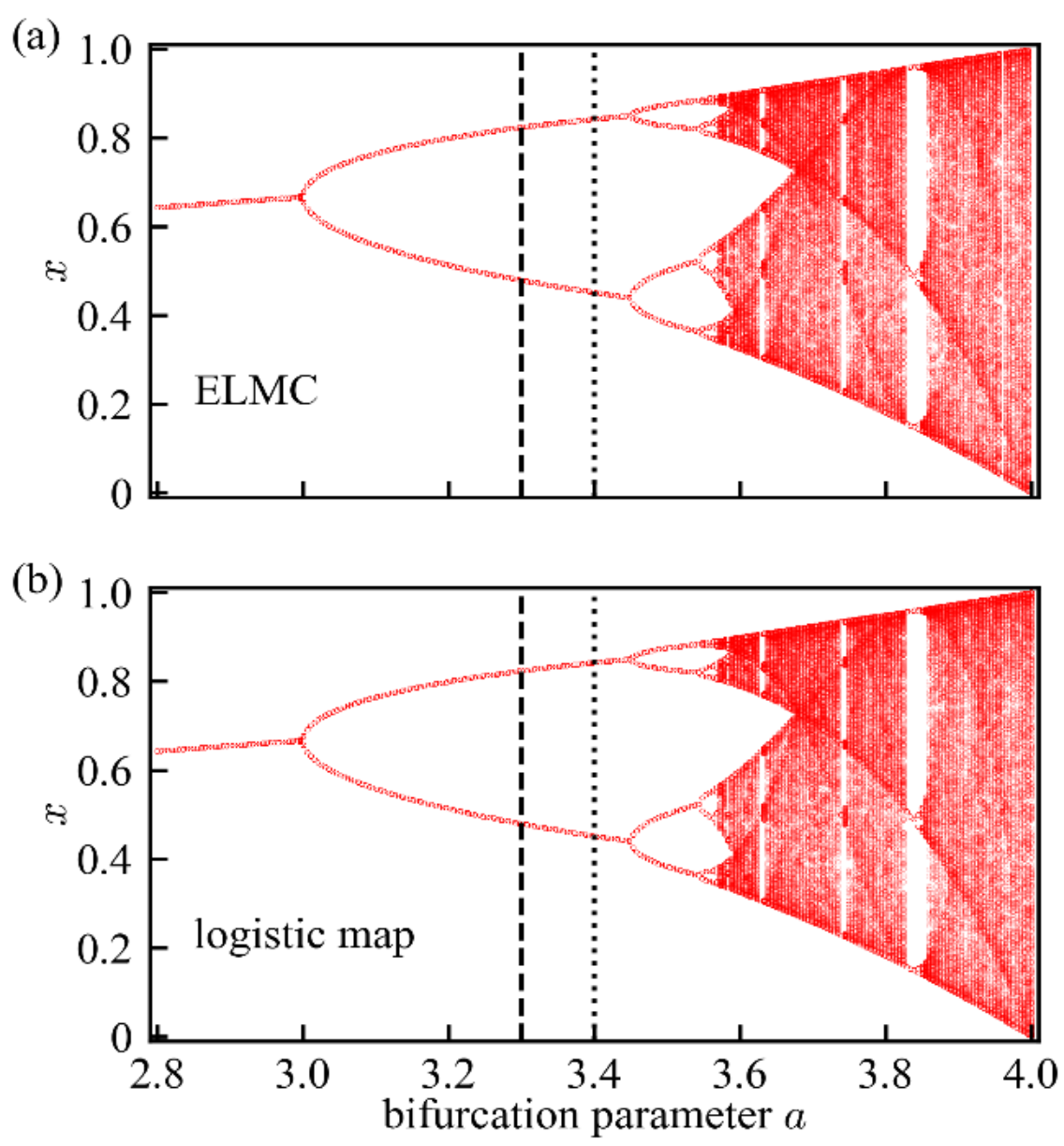

Fig. 4 (a) Inferred bifurcation diagram by ELMC of the logistic map system. (b) Original bifurcation diagram. The dashed and dotted vertical lines indicate the parameter values at which training of ELMC was performed.

structure, including period-doubling bifurcations converging to chaos and periodic windows converging to chaos.

Thus, ELMC exhibits a remarkable ability to reproduce the dynamics at bifurcation parameter values significantly distant from those used during the training phase.

## B. The case of Hénon map

Subsequently, we extended our analysis to two-dimensional discrete-time dynamical systems, using a typical two-dimensional chaotic Hénon map [13], which is defined by the following equation:

$$\begin{cases} x(n+1) = 1 - ax(n)^2 + y(n) \\ y(n+1) = bx(n) \end{cases} \tag{14}$$

In this context, parameter $a$ is considered a bifurcation parameter, whereas parameter $b$ is typically fixed at 0.3. Two distinct sets of bifurcation parameters, $a_1 = 0.95$ ($c_1 = 0$) and $a_2 = 1.0$ ($c_2 = 1$), were used for the training. For each parameter set, we generated $N_{init} = 100$ trajectories with a length of $N = 4$ starting from initial values of $x_1$ and $x_2$, randomly chosen from the interval $[-0.1, 0.1]$. A training dataset was constructed using these trajectories. The scatter plots of the data points of the trajectories in the training dataset are shown in Fig. 5. Using this training dataset, we trained ELMC with the following parameters: $N_{in}$=2, $N_h$=50, $N_c$=1, $N_{out}$=2, α=0.2, and β=0.00005.

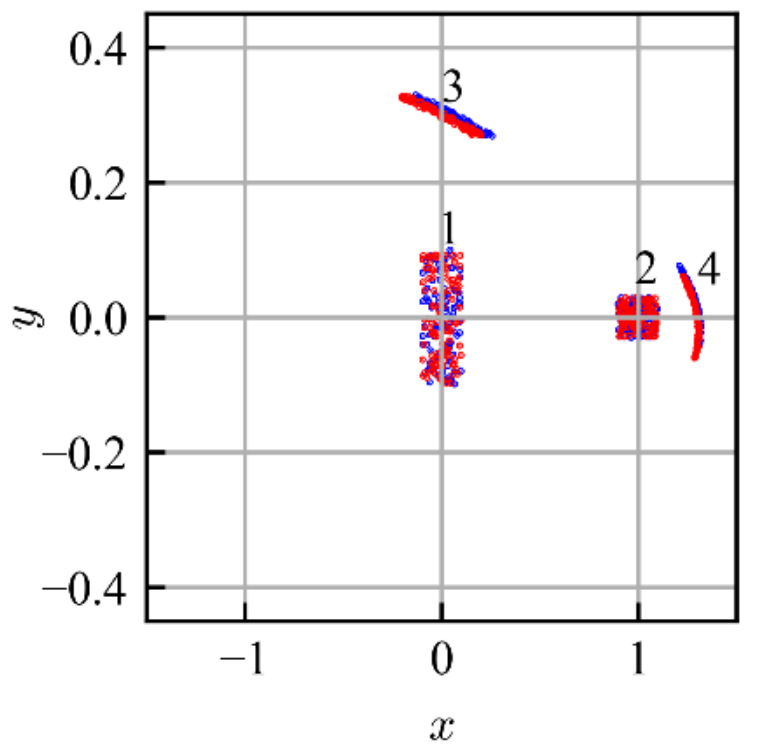

Fig. 5 Scattered data points $\left(x_i^j(n), y_i^j(n)\right)$ ($j = 1 \cdots N_{init}$, $n = 1, 2, 3, 4, i = 1, 2$) for trajectories in the training dataset. Each cluster of data points is accompanied by a number denoting time step $n$. The red and blue data points represent $\left(x_1^j(n), y_1^j(n)\right)$ for $a = a_1$ and $\left(x_2^j(n), y_2^j(n)\right)$ for $a = a_2$, respectively.

Fig. 6 (a) shows the time series of $x(n)$ for the original Hénon map at a chaotic bifurcation parameter value of $a = 1.4$ together with the time series generated by ELMC for $c = 9$ corresponding to $a = 1.4$. The results indicate that while ELMC achieves accurate short-term predictions, it does not maintain this accuracy for long-term predictions. However, ELMC effectively replicates the chaotic attractor (Hénon attractor), as shown in Fig. 6 (b) and (c).

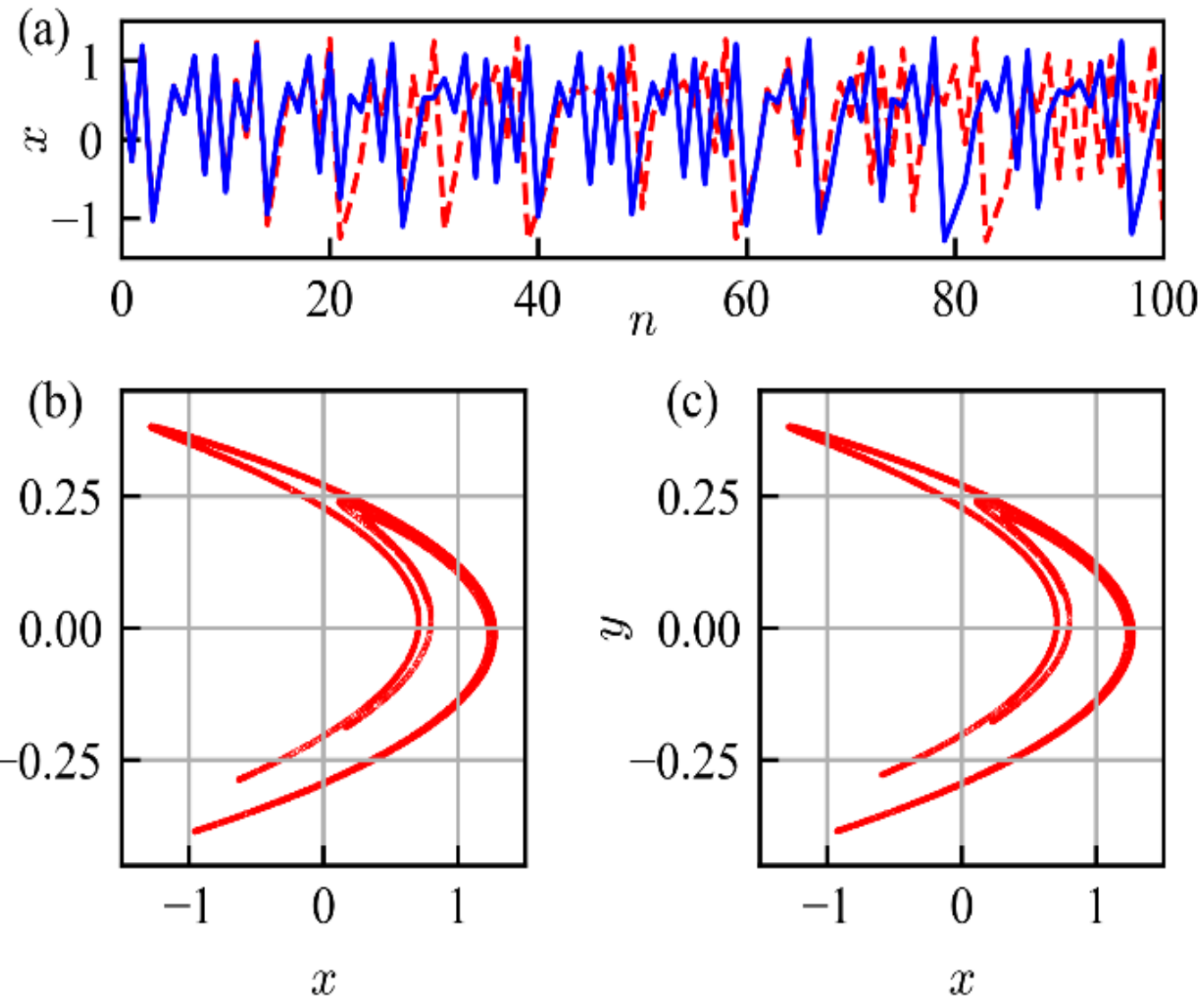

Fig. 6 (a) Time-series $\{x(n)\}$ of the original Henon map (red dashed curve) and its trained counterpart for a dynamical system (blue solid curve) for $a = 1.4$ ($c = 9$); (b), (c) Attractor of the trained dynamical system with $c = 9$ and the Hénon attractor corresponding to $a = 1.4$, respectively. Plots are generated from trajectories starting with random initial points, excluding the first 100 steps as transient from original trajectories with a length of 10000 steps.

In the case of the Henon map, as with the logistic map, the ELMC successfully reproduced the dynamical behaviors even at distant bifurcation parameter values from those used in the training phase.

## V. MECHANISM FOR INTERPOLATION AND EXTRAPOLATION OF DYNAMICAL SYSTEMS

In this section, we elucidate the mechanism for reproducing the dynamics at unlearned bifurcation parameter values by adjusting the control input values.

From the learning procedure, it is evident that ELMC learns numerous one-step transition processes $\boldsymbol{u}(n) \to \boldsymbol{u}(n+1)$ which are observed in the time series of underlying dynamical system, $\{\boldsymbol{u}(0), \boldsymbol{u}(1), \boldsymbol{u}(2), \cdots\}$; however, it does not learn a sequence of these one-step transition processes. In essence, the learning process of the time-series data produced by a given dynamical system $\boldsymbol{u}(n+1) = f\big(a, \boldsymbol{u}(n)\big)$ with ELMC entails acquiring knowledge of the input-output relationship intrinsic to the function $f(a, \boldsymbol{u})$, thereby creating an approximate function for $f(a, \boldsymbol{u})$. Function $f^{ELMC}(c, \boldsymbol{u})$ defined by Eq. (11) at $c = c_1$ and $c_2$ serves as an approximation of $f(a_1, \boldsymbol{u})$ and $f(a_2, \boldsymbol{u})$, respectively. Fig. 7 shows that function $f^{ELMC}(c, x)$ constructed for the logistic map serves as an approximation of $f(a, x)$. The figures show the success of the approximation for both learned parameters (shown in (a) and (b)) and unlearned parameters (shown in (c) and (d)).

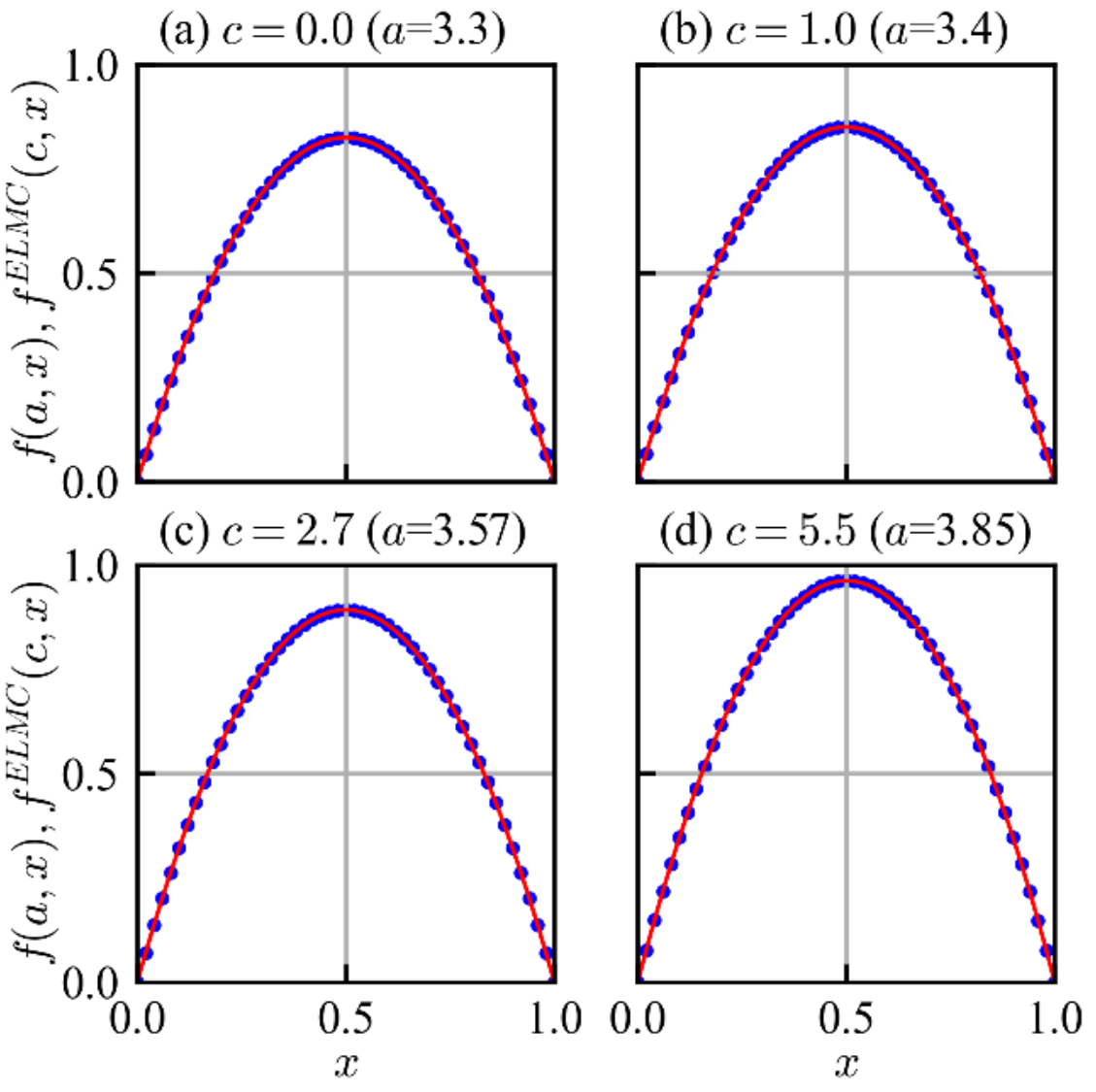

Fig. 7 Function $y = f^{ELMC}(c, x)$ constructed for the logistic map (blue dots) and $y = ax(1-x)$ (red curve) with $a = a(c), as\ expressed$ in Eq.(13) .

A key point is the relationship between the bifurcation parameter $a$ and control input $c$, expressed in Eq. (13), which was successfully learned. We employed the power expansion of the hyperbolic tangent function to investigate the behaviors of the hidden layer neurons associated with the control input $c$:

$$\tanh(\xi+\delta)=\tanh(\xi)+\delta\,\mathrm{sech}^2(\xi)+O(\delta^2), \tag{15}$$

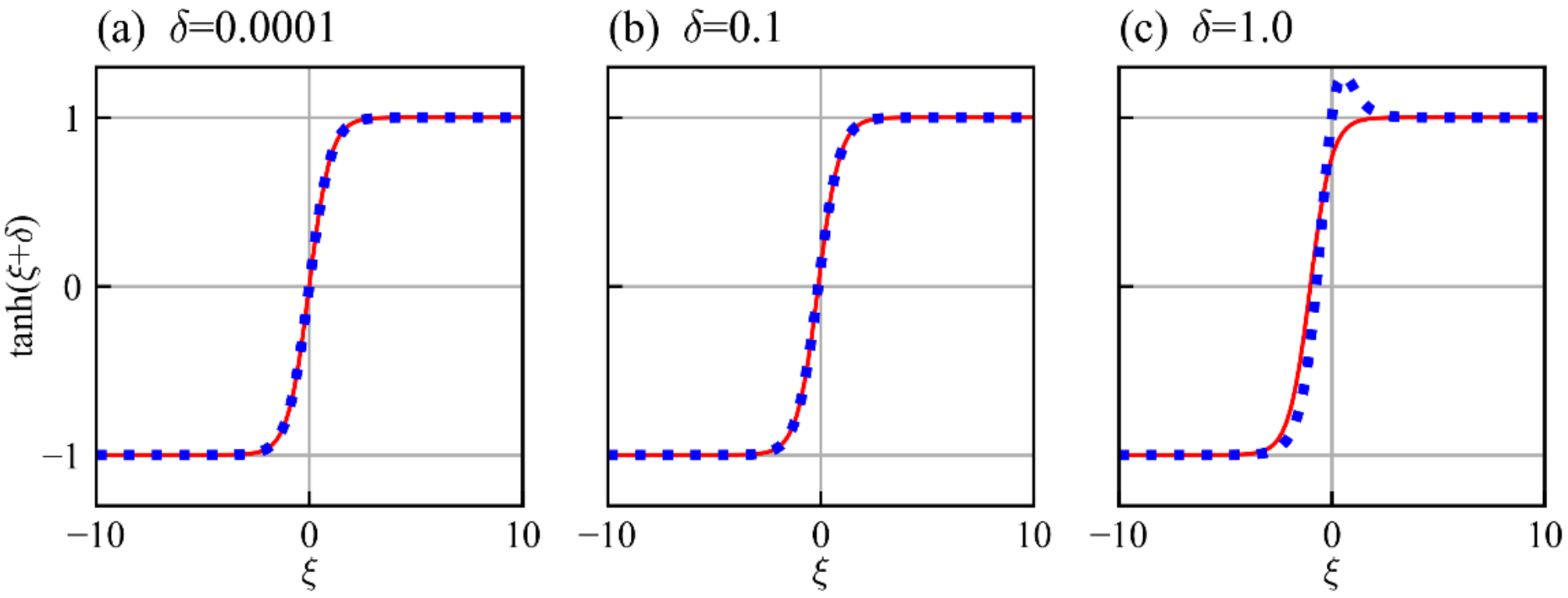

Fig. 8 Graph of function $\tanh(\xi+\delta)$ (solid curve) and its first-order approximation with respect to $\delta$ (dotted curve). (a) $\delta=0.0001$; (b) $\delta=0.1$; (c) $\delta=1.0$.

The effects of this first-order approximation are illustrated in Fig. 8. In this figure, the case of $\delta<0$ is not explicitly shown, but such a case can be estimated using $\tanh(\xi-\delta)=-\tanh(-\xi+\delta)$. It is evident that this approximation is quite successful for $|\delta|\leq 0.1$. The expansion of the activation function of the hidden layer neurons with respect to $c-c_1$ up to the first order is obtained as follows:

$$\tanh\left(\alpha\boldsymbol{W}^{in}\boldsymbol{u}+\beta\boldsymbol{W}^{c}c+\boldsymbol{\theta}\right)\approx\tanh\left(\alpha\boldsymbol{W}^{in}\boldsymbol{u}+\beta\boldsymbol{W}^{c}c_1+\boldsymbol{\theta}\right)+(c-c_1)\boldsymbol{W}^{c}\beta\,\mathrm{sech}^2\left(\alpha\boldsymbol{W}^{in}\boldsymbol{u}+\beta\boldsymbol{W}^{c}c_1+\boldsymbol{\theta}\right). \tag{16}$$

To evaluate the effectiveness of this approximation in a practical sense requires the estimation of δ, which corresponds to $\beta\boldsymbol{W}^{c}(c-c_1)$, where elements of $\boldsymbol{W}^{c}$ are of order $O(1)$. We estimate δ in the ELMC for the logistic map as well as the Hénon map. For the logistic map, we used $\beta=0.00002$, with $c-c_1$ varying from -5.0 (corresponding to $a=2.8$) to 7.0 (corresponding to $a=4.0$). Consequently, the maximum value of $|\delta|$ was constrained up to the order of $10^{-4}$. In the analysis of the strange attractor of the Hénon map shown in Fig. 6, we set $\beta=0.00005$ and $c-c_1=9$, resulting in $\delta\sim 0.00045$. In both cases, the value of δ falls within the range where the approximations obtained using Eqs. (15) and (16) are valid.

By substituting the expanded terms described by Eq. (16) into $f^{ELMC}(c,\boldsymbol{u})$, the following equation is derived:

$$\begin{aligned} \boldsymbol{f}^{ELMC}(c,\boldsymbol{u}) &\approx \boldsymbol{F_0}(\boldsymbol{u}) + (c - c_1)\boldsymbol{F_1}(\boldsymbol{u}) \\ \boldsymbol{F_0}(\boldsymbol{u}) &= \boldsymbol{W}^{out}\mathbf{tanh}\left(\alpha\boldsymbol{W}^{in}\boldsymbol{u} + \beta\boldsymbol{W}^{c}c_1 + \boldsymbol{\theta}\right) \\ \boldsymbol{F_1}(\boldsymbol{u}) &= \boldsymbol{W}^{out}\boldsymbol{W}^{c}\beta\,\mathbf{sech}^{2}\left(\alpha\boldsymbol{W}^{in}\boldsymbol{u} + \beta\boldsymbol{W}^{c}c_1 + \boldsymbol{\theta}\right). \end{aligned} \tag{17}$$

Setting $c = c_1$ and $c_2$ in the first equation of Eq. (17), we obtain

$$\begin{aligned} \boldsymbol{f}^{ELMC}(c_1,\boldsymbol{u}) &\approx \boldsymbol{F}_0(\boldsymbol{u}) \\ \boldsymbol{f}^{ELMC}(c_2,\boldsymbol{u}) &\approx \boldsymbol{F}_0(\boldsymbol{u}) + (c_2 - c_1)\boldsymbol{F}_1(\boldsymbol{u}). \end{aligned} \tag{18}$$

Assuming $\boldsymbol{f}^{ELMC}(c,\boldsymbol{u})$ at $c = c_1$ and $c_2$ can approximate $\boldsymbol{f}(a_1,\boldsymbol{u})$ and $\boldsymbol{f}(a_2,\boldsymbol{u})$ of the function $\boldsymbol{f}(a,\boldsymbol{u})$ for the target dynamical system, respectively, $\boldsymbol{f}^{ELMC}(c_1,\boldsymbol{u}) \approx \boldsymbol{f}(a_1,\boldsymbol{u})$ and $\boldsymbol{f}^{ELMC}(c_2,\boldsymbol{u}) \approx \boldsymbol{f}(a_2,\boldsymbol{u})$, and Eq. (18) can be rewritten as:

$$\begin{aligned} \boldsymbol{f}(a_1,\boldsymbol{u}) &\approx \boldsymbol{F}_0(\boldsymbol{u}) \\ \boldsymbol{f}(a_2,\boldsymbol{u}) &\approx \boldsymbol{F}_0(\boldsymbol{u}) + (c_2 - c_1)\boldsymbol{F}_1(\boldsymbol{u}). \end{aligned} \tag{19}$$

From these equations, the expressions for $\boldsymbol{F}_0(\boldsymbol{u})$ and $\boldsymbol{F}_1(\boldsymbol{u})$ in terms of $\boldsymbol{f}(a_1,\boldsymbol{u})$ and $\boldsymbol{f}(a_2,\boldsymbol{u})$ can be derived as follows:

$$\begin{aligned} \boldsymbol{F}_0(\boldsymbol{u}) &\approx \boldsymbol{f}(a_1,\boldsymbol{u}) \\ \boldsymbol{F}_1(\boldsymbol{u}) &\approx \left(\boldsymbol{f}(a_2,\boldsymbol{u}) - \boldsymbol{f}(a_1,\boldsymbol{u})\right)/(c_2 - c_1)\,. \end{aligned} \tag{20}$$

Using these expressions, we derive the final expression for $\boldsymbol{f}^{ELMC}(c,\boldsymbol{u})$ as follows:

$$\boldsymbol{f}^{ELMC}(c,\boldsymbol{u}) \approx \boldsymbol{f}(a_1,\boldsymbol{u}) + \frac{\boldsymbol{f}(a_2,\boldsymbol{u}) - \boldsymbol{f}(a_1,\boldsymbol{u})}{c_2 - c_1}(c - c_1) \tag{21}$$

Eq. (21) implies that $\boldsymbol{f}^{ELMC}(c,\boldsymbol{u})$ embodies a linear interpolation scheme that can derive a function $\boldsymbol{f}(a,\boldsymbol{u})$ for any control input $c$ by realizing the change in function $\boldsymbol{f}(a,\boldsymbol{u})$ corresponding to different values of the control input $c$. However, at this stage, the correspondence between $\boldsymbol{f}(a,\boldsymbol{u})$ and $\boldsymbol{f}^{ELMC}(c,\boldsymbol{u})$ remains unclear because the relationship between $a$ and $c$ has not yet been determined.

Now, let us consider the case in which the bifurcation parameter $a$ is linearly incorporated into $\boldsymbol{f}(a,\boldsymbol{u})$ and expressed as

$$\boldsymbol{f}(a,\boldsymbol{u}) = a\boldsymbol{p}(\boldsymbol{u}) + \boldsymbol{q}(\boldsymbol{u}) \tag{22}$$

where, $\boldsymbol{p}(\boldsymbol{u})$ denotes the term for which the bifurcation parameter $a$ acts as a multiplicative term, and $\boldsymbol{q}(\boldsymbol{u})$ is an additional term, which is free of the bifurcation parameter. Inserting Eq. (22) into the $f(a_i, \boldsymbol{u})$ in Eq. (21), $f^{ELMC}(c, \boldsymbol{u})$ can be expressed as:

$$\begin{aligned} \boldsymbol{f}^{ELMC}(c, \boldsymbol{u}) &\approx a_1 \boldsymbol{p}(\boldsymbol{u}) + \boldsymbol{q}(\boldsymbol{u}) + \frac{a_2 - a_1}{c_2 - c_1}(c - c_1)\boldsymbol{p}(\boldsymbol{u}) \\ &= \left(a_1 + \frac{a_2 - a_1}{c_2 - c_1}(c - c_1)\right)\boldsymbol{p}(\boldsymbol{u}) + \boldsymbol{q}(\boldsymbol{u}) \end{aligned} \tag{23}$$

The function $a(c)$ in Eq. (13) is the same as the coefficient of $\boldsymbol{p}(\boldsymbol{u})$. Finally, $f^{ELMC}(c, \boldsymbol{u})$ is formulated as

$$\boldsymbol{f}^{ELMC}(c, \boldsymbol{u}) \approx a(c)\boldsymbol{p}(\boldsymbol{u}) + \boldsymbol{q}(\boldsymbol{u}) \tag{24}$$

The logistic map has the structure represented by Eq. (22), and the functions $p(x)$ and $q(x)$ are expressed as follows:

$$\begin{aligned} p(x) &= x(1 - x) \\ q(x) &= 0 \end{aligned} \tag{25}$$

Thus, $f^{ELMC}(c, x)$ for the logistic map can be rewritten as:

$$f^{ELMC}(c, x) \approx a(c)x(1 - x) \tag{26}$$

Similarly, the Hénon map is expressed as in Eq. (27), where $\boldsymbol{p}(\boldsymbol{x})$ and $\boldsymbol{q}(\boldsymbol{x})$ are vector functions of vector $\boldsymbol{x} = (x, y)^T$, expressed as

$$\begin{aligned} \boldsymbol{p}\begin{pmatrix} x \\ y \end{pmatrix} &= \begin{pmatrix} x^2 \\ 0 \end{pmatrix} \\ \boldsymbol{q}\begin{pmatrix} x \\ y \end{pmatrix} &= \begin{pmatrix} 1 + y \\ bx \end{pmatrix} \end{aligned} \tag{27}$$

Thus, for the Hénon map, $\boldsymbol{f}^{ELMC}(c, \boldsymbol{x})$ is expressed as:

$$\boldsymbol{f}^{ELMC}(c, \boldsymbol{x}) \approx \begin{pmatrix} 1 - a(c)x^2 + y \\ bx \end{pmatrix} \tag{28}$$

In summary, we obtained the following results: (1) First, if the hyperbolic tangent function for the activation function of the hidden layer neurons of the ELMC can be well approximated by a first-order expansion with respect to the control input, the function $\boldsymbol{f}(a,\boldsymbol{u})$, provides a dynamical rule for different values of the bifurcation parameter $a$ to be delineated through linear interpolation (and also extrapolation) of the control input $c$ (Eq. (21)). The relationship between the $\boldsymbol{f}^{ELMC}(c,\boldsymbol{u})$ obtained using Eq. (21) at a given value of $c$ and $\boldsymbol{f}(a,\boldsymbol{u})$ at a specific value of the bifurcation parameter $a$ remains undetermined at this stage, except for the equalities $\boldsymbol{f}^{ELMC}(c=0,\boldsymbol{u})=\boldsymbol{f}(a_1,\boldsymbol{u})$ and $\boldsymbol{f}^{ELMC}(c=1,\boldsymbol{u})=\boldsymbol{f}(a_2,\boldsymbol{u})$. (2) If function $\boldsymbol{f}(a,\boldsymbol{u})$ includes the bifurcation parameter $a$ in a linear form, an approximate function can be expressed via a linear interpolation and extrapolation scheme of the control parameter $c$. Consequently, by manipulating $c$, $\boldsymbol{f}(a,\boldsymbol{u})$ $can\ be\ reproduced$ at any $a$.

## VI. REALIZATION OF ELMC IN TERMS OF CONTINUOUS-TIME DYNAMICAL SYSTEMS

### A. Case of Lorenz system

The question is whether ELMC is capable of forecasting the bifurcation structures in continuous-time dynamical systems, that is, vector fields, as well as each dynamical behavior. We attempted to answer this question by applying ELMC to the Lorenz system [14] as a typical example, which is described by the following differential equations:

$$\begin{aligned}\frac{d\boldsymbol{x}}{dt}&=\boldsymbol{F}(\boldsymbol{x})\\ \boldsymbol{F}(\boldsymbol{x})&=\begin{pmatrix}p(y-x)\\ x(r-z)-y\\ xy-bz\end{pmatrix}\end{aligned}\tag{29}$$

where $\boldsymbol{x}=(x,y,z)^T$. In the present study, the value of parameter $r$ was varied while keeping other parameters fixed at $p=10$ and $b=8/3$, as has been widely studied. To integrate these equations numerically, the 4-th order Runge-Kutta method (RK4) was used.:

$$\begin{aligned}&\boldsymbol{x}(t+\Delta t)=\boldsymbol{G}_{RK}\big(\boldsymbol{x}(t)\big)=\boldsymbol{x}(t)+\frac{\Delta t}{6}(\boldsymbol{k}_1+2\boldsymbol{k}_2+2\boldsymbol{k}_3+\boldsymbol{k}_4)\\ &\boldsymbol{k}_1=\boldsymbol{F}\big(\boldsymbol{x}(t)\big)\\ &\boldsymbol{k}_2=\boldsymbol{F}(\boldsymbol{x}(t)+\Delta t\boldsymbol{k}_1/2)\\ &\boldsymbol{k}_3=\boldsymbol{F}(\boldsymbol{x}(t)+\Delta t\boldsymbol{k}_2/2)\\ &\boldsymbol{k}_4=\boldsymbol{F}(\boldsymbol{x}(t)+\Delta t\boldsymbol{k}_3)\end{aligned}\tag{30}$$

where $\Delta t$ represents the time step of the numerical simulation and was set to $\Delta \mathrm{t}=0.001$.

Although this system consists of coupled-differential equations, the system can be viewed as a map under the RK4 numerical approximation, namely a discrete-time dynamical system, such as $\boldsymbol{x}(n+1)=\boldsymbol{G}_{RK}\big(\boldsymbol{x}(n)\big)$, rewriting $\boldsymbol{x}(n\Delta t)$ as $\boldsymbol{x}(n)$, where the function $\boldsymbol{G}_{RK}$ is defined in Eq.(30). The function $\boldsymbol{G}_{RK}(\boldsymbol{x})$ includes terms up to the 11-th order of $\Delta t$ and up to the 3rd order of $r$. For truncation up to the second order of $\Delta t$, $\boldsymbol{G}_{RK}$ retains the terms up to first order of $r$. Given $\Delta t = 0.001$, higher-order terms of $\Delta t$ in $G_{RK}$ can be negligible. Therefore, it is reasonable to infer that $\boldsymbol{G}_{RK}$ effectively includes only terms that do not exceed the second order of $r$, thus possessing the structure represented by Eq. (22).

The training dataset was constructed according to the method explained in Section III using, for example, the parameter values $r_1 = 23$, $r_2 = 24$, $c_1 = 0$, $c_2 = 1$, $N_{init} = 10$ and $N = 5000$ ($t = 0 \sim 5$). The $N_{init}$ distinct initial values of $\boldsymbol{x}(0)$ were randomly chosen in a neighborhood of the two fixed points $\left(\pm\sqrt{b(r-1)}, \pm\sqrt{b(r-1)}, r-1\right)$ for each of $r = r_1$ and $r = r_2$. Fig. 9 shows all the overlaid training data trajectories. All trajectories eventually converge to one of the fixed points. Using this dataset, we trained ELMC with parameters $N_x = 3$, $N_c = 1$, $N_h = 200$, $\alpha = 0.01$, and $\beta = 0.0001$.

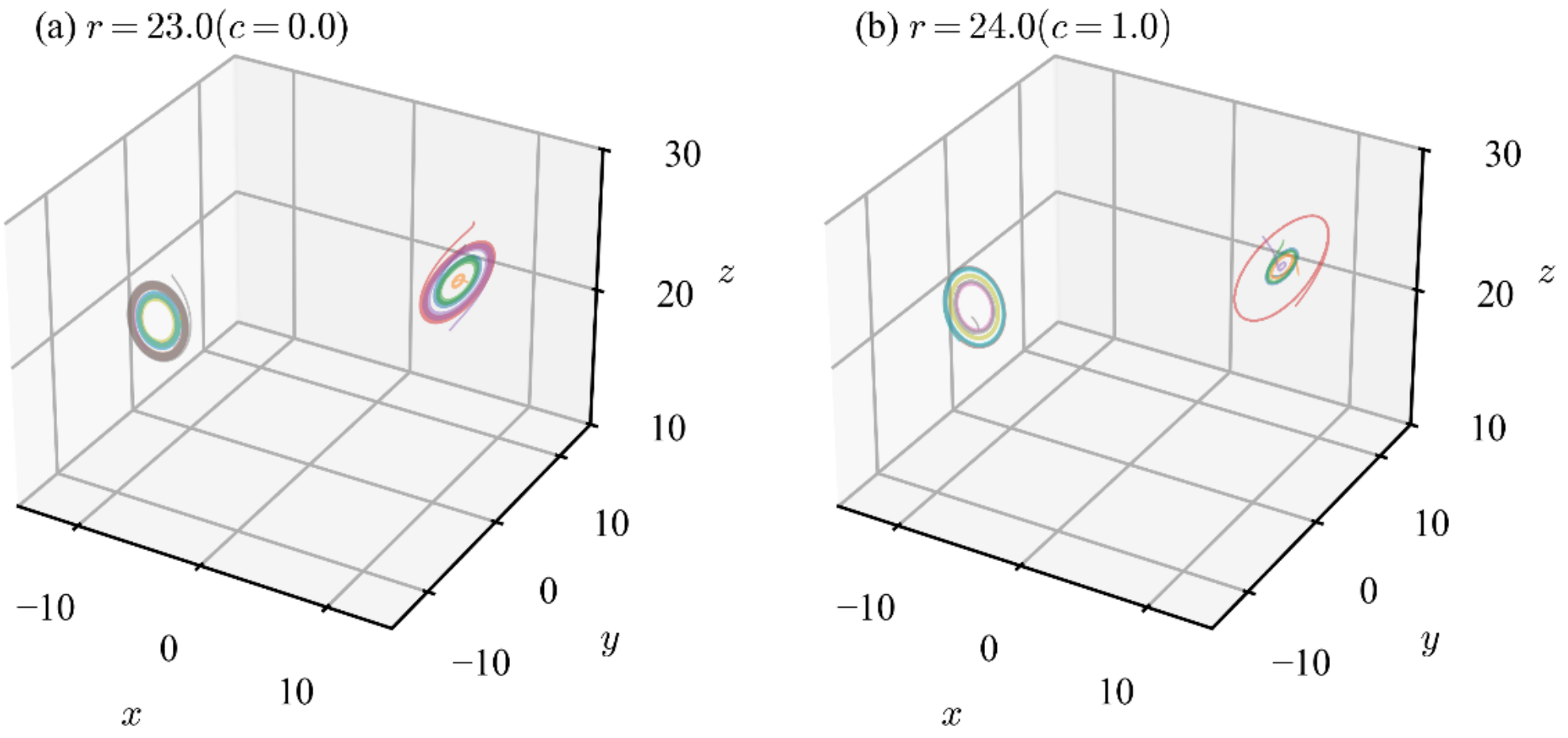

Fig. 9 Trajectories within the training dataset. (a) $r = 23$. (b) $r = 24$.

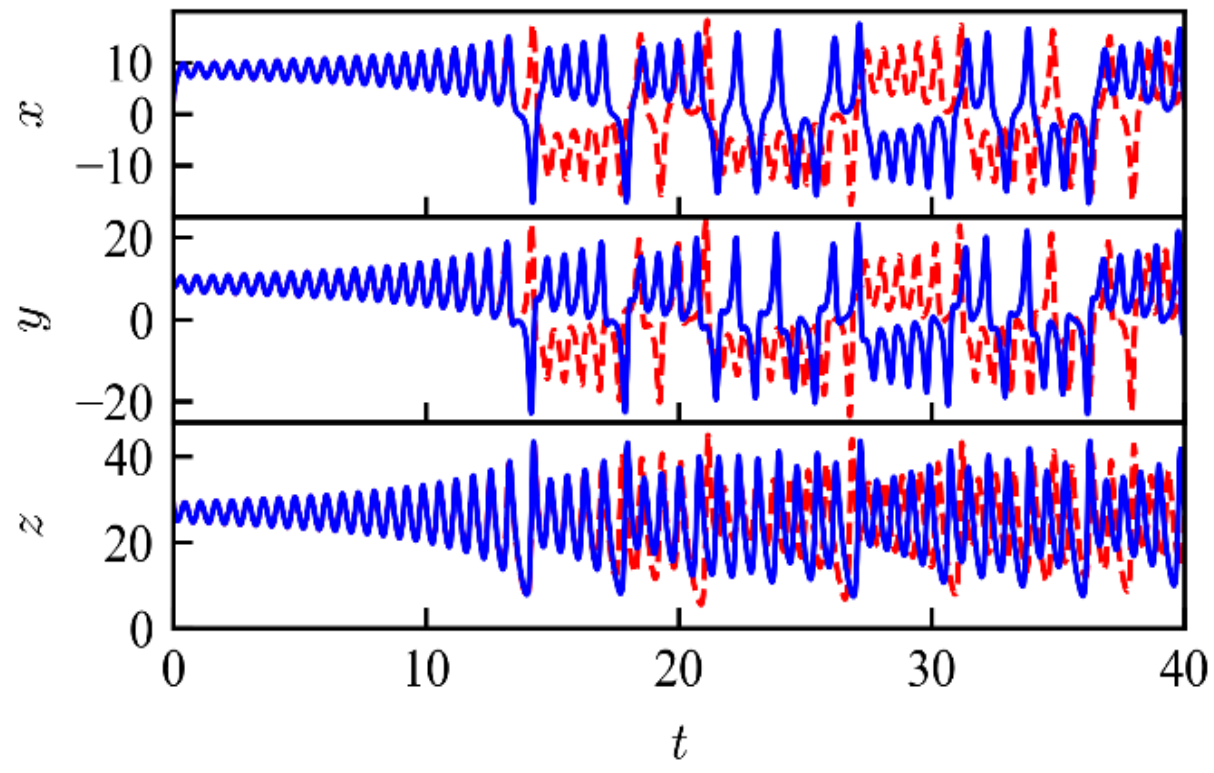

Fig. 10 Comparison of predicted (blue solid curve) and actual (red dashed curve) time-series for the $x$, $y$, and $z$ variables of the Lorenz system at $r = 28$, starting from the same initial value of $(x, y, z)$. The predicted time-series was generated by the ELMC trained employing a control input of $c = 5$.

Fig. 10 illustrates the predicted time series of variables of $x_1$, $x_2$, and $x_3$ in the Lorenz system at $r = 28$, which were generated using the trained ELMC with control input $c = 5$. Notably, the predicted time series is closely aligned with the actual time series up to $t = 10$. Although precise predictions are no longer feasible beyond this time point, the overall structure of the strange attractor appears to have been reproduced throughout the prediction range. To confirm this observation, we studied the trajectories in phase space using the Lorenz plot. The results are shown in Fig. 11. Panel (a) compares the phase space of the actual trajectories of the strange attractor at $r = 28$ with those predicted by ELMC. This reveals that ELMC can reproduce the structure of a strange attractor. The Lorenz plots shown in Panel 11(b) demonstrate that ELMC reflects the deterministic law governing the chaotic trajectories, which are embedded into Lorenz chaos at r = 28.

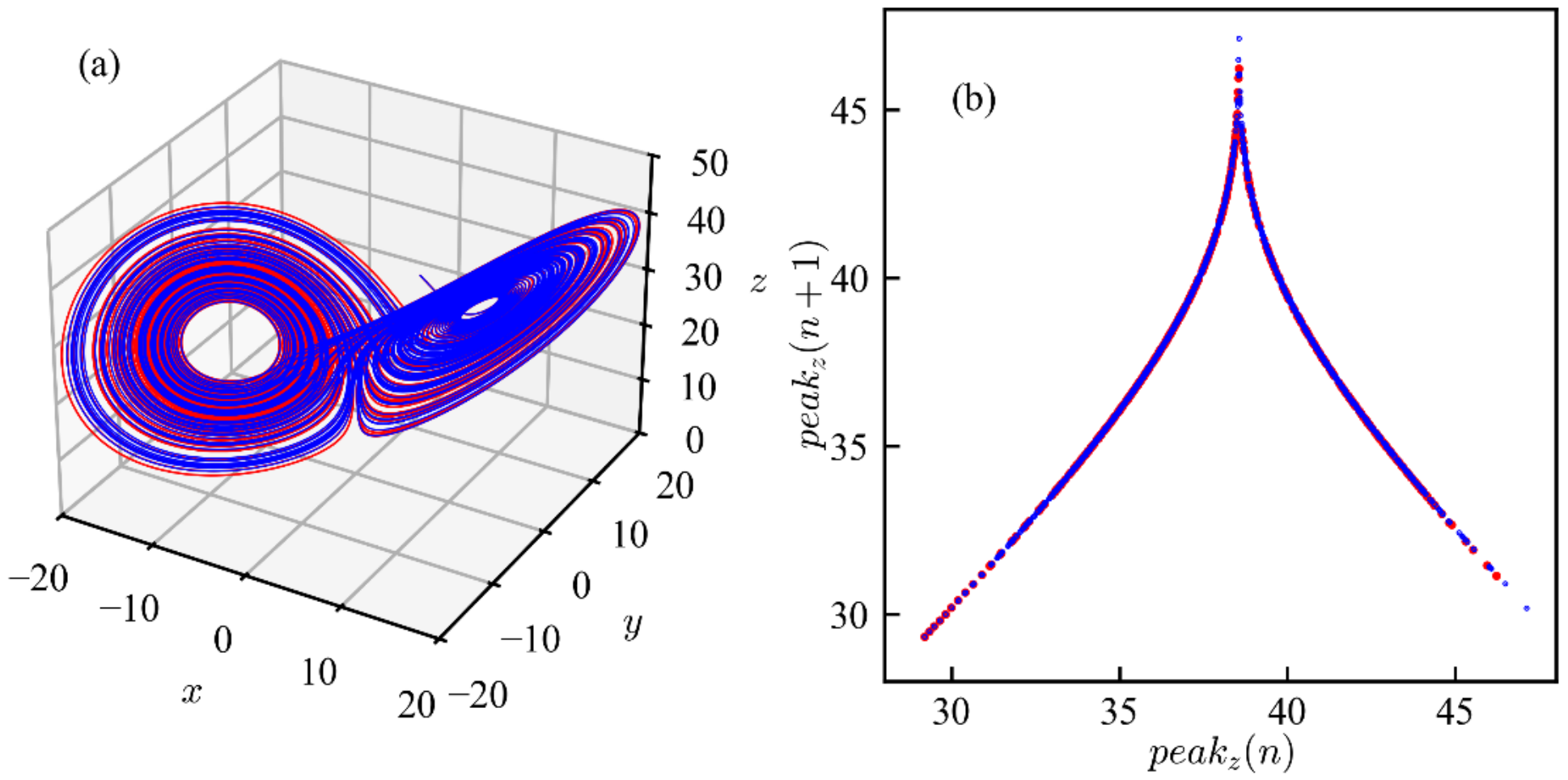

Fig. 11 Trajectory of a strange attractor at $r = 28$ (red curve) and its predicted trajectory by ELMC for $c = 5$ (blue curve). (a) Attractor represented in three-dimension $(x, y, z)$; (b) Lorenz plot obtained from local peaks of the time-series of the third variable, $z$.

In the subsequent analysis, we examined the ability of the ELMC to replicate the bifurcation structure. A bifurcation diagram of the Lorenz system for the region of bifurcation parameter $r = 15 - 35$ is presented in Fig. 12. As illustrated in the figure, the ELMC reproduces the overall bifurcation

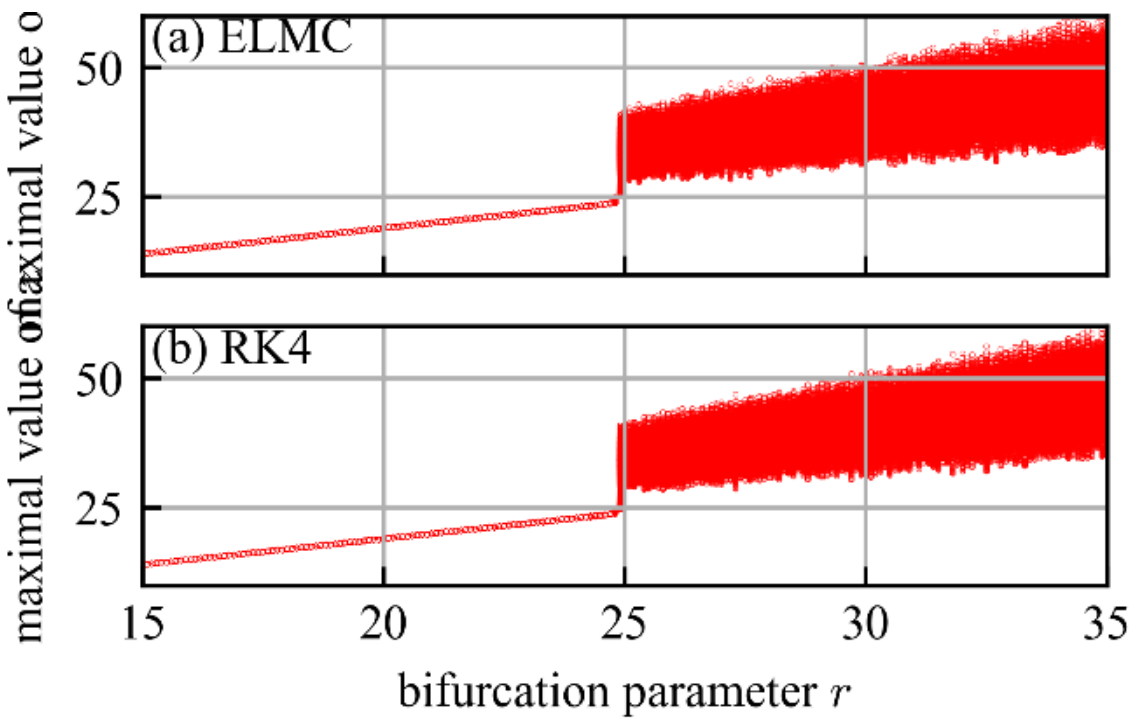

Fig. 12 Bifurcation diagrams of the Lorenz system. (a) Predicted by ELMC. (b) Obtained by the original equations using the fourth order Runge-Kutta method. The diagrams are plotted using the local maximum values of $\{z(t)\}$.

structure well, including the bifurcation point from a fixed point to chaos.

The validity of the approximations Eq. (15), that is, Eq. (16) were verified in the above analysis. Given the parameter values $\beta = 0.0001$ and $c_1 = 0$, we have $|\delta| = |\beta \boldsymbol{W}^c(c - c_1)| \sim 0.0001|c|$. The range of the bifurcation parameter $r$ in the aforementioned analysis is $15 - 35$, corresponding to the range of $c\ from - 8\ to\ 12$. Therefore, the maximum value of $|c| < 12$ implies that $max(|\delta|) < 0.0012$. Consequently, the approximations to Eq. (15) (i.e. Eq. (16)) are valid for this range (Fig. 8).

Kim et al. deduced the period doubling bifurcation diagram of the Lorenz system for the bifurcation parameter range $r = 99.5$ to $r = 100.5$ using the echo state network with a control input. This network was trained using eight single-cycle trajectories with $r$ values close to 100. Let us conduct a similar inference using ELMC. We created a training dataset consisting of 50 transient trajectories for each parameter setting $r = 100.2$ and $r = 100.3$ associated with $c = 0$ and $c = 1$, respectively, which eventually converged to a single-cycle orbit.

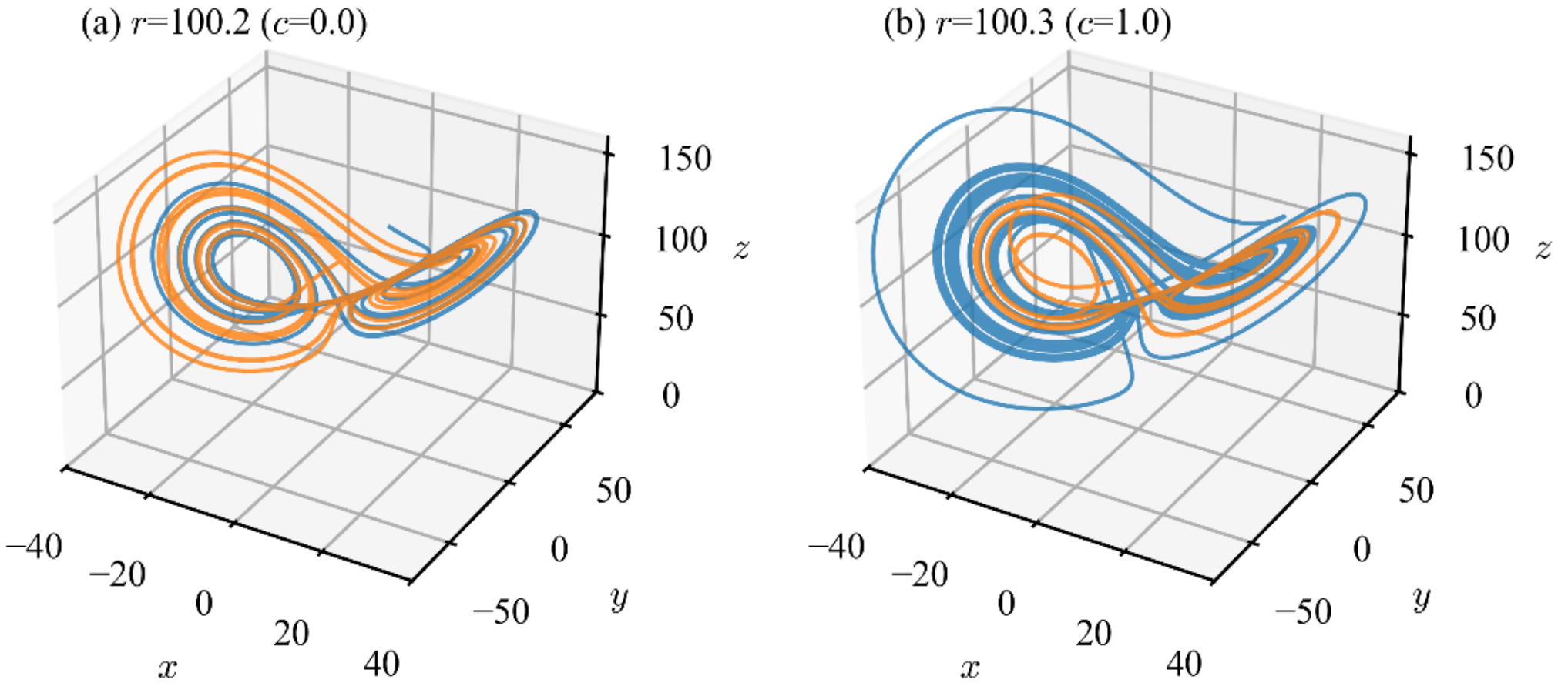

Fig. 13 Two representative trajectories in the training dataset consisting of transient trajectories for each parameter setting at (a) $r = 100.2$ and (b) $r = 100.3$ associated with $c = 0$ and $c = 1$, respectively..

Using this training dataset, we trained ELMC with $N_h = 200$, $\alpha = 0.002$, and $\beta = 0.00006$. Fig. 13 shows two representative trajectories at $r = 100.2$ (left) and $r = 100.3$ (right) for the training dataset. The bifurcation diagram predicted by the trained ELMC is presented in Fig. 14. The figure illustrates that the trained ELMC effectively replicates the bifurcation structures, calculated using the fourth-order Runge-Kutta method. With the present ELMC, $n$-cycle orbits were accurately predicted at an early period, but after a certain period, the phase deviated while maintaining amplitude and shape for some unknown reason.

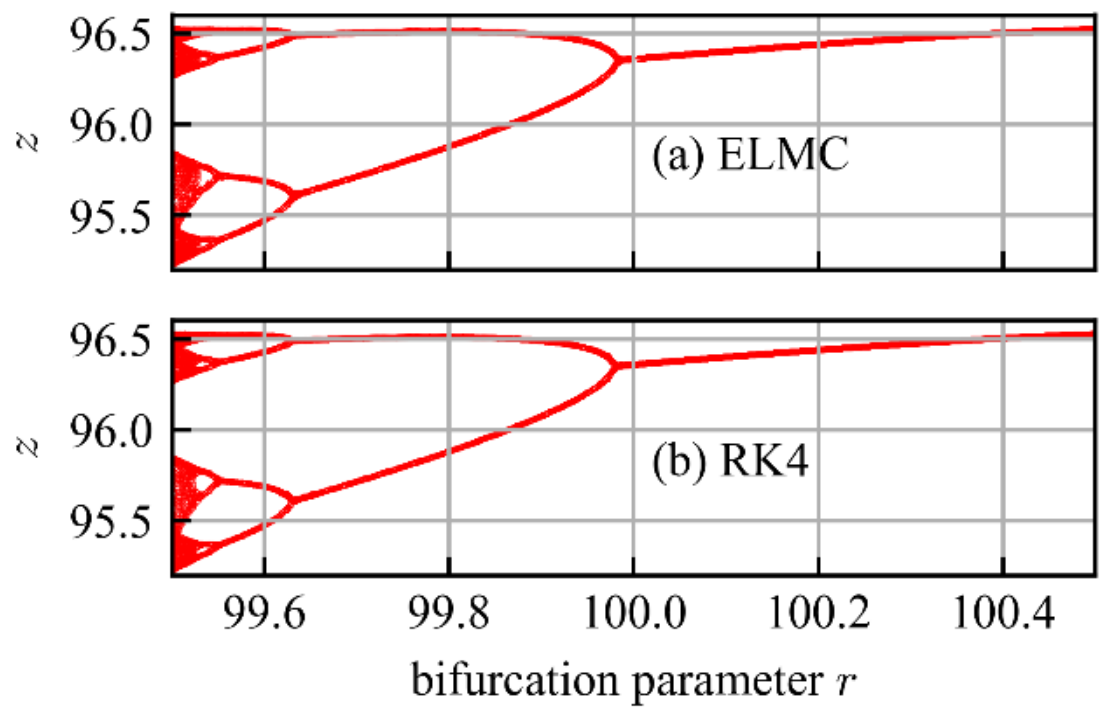

Fig. 14 Bifurcation diagram around $r \approx 100$ for the Lorenz system, constructed based on the trajectories on the Poincaré section defined by the plane $z = 0$ and $\dot{z} > 0$ for (a) ELMC and (b) the fourth-order Runge-Kutta method.

## B. Case of Rössler system

The Rössler system is described by the following differential equations:

$$\frac{d\boldsymbol{x}}{dt} = \begin{pmatrix} -y - z \\ x + ay \\ b + xz - \gamma z \end{pmatrix} \tag{31}$$

Similar to the method used in the Lorenz system, the fourth-order Runge-Kutta method was employed for the numerical integration of these equations. In this case, a step size of $\Delta t = 0.01$ was used, varying the value of the parameter $b$, while keeping the other parameters fixed at $a = 0.2$ and $\gamma = 5.7$. When truncating up to the third order of $\Delta t$, the $\boldsymbol{G}_{RK}$ in Eq. (30) includes terms up to the first order of $b$ and has the structure represented by Eq. (22)

For the bifurcation parameters $b = 8$ and $b = 6$, we generated trajectories with a time span of 10, starting with 20 initial conditions, which were randomly selected within the region defined by $x, y \in [-5, 5]$ and $z \in [0, 4]$ for the training dataset. Fig. 15 shows these trajectories, each eventually converging to a fixed point corresponding to the respective bifurcation parameter values. A training dataset was generated using these trajectories, associating $b = 8$ with $c = 0$ and $b = 6$ with $c = 1$.

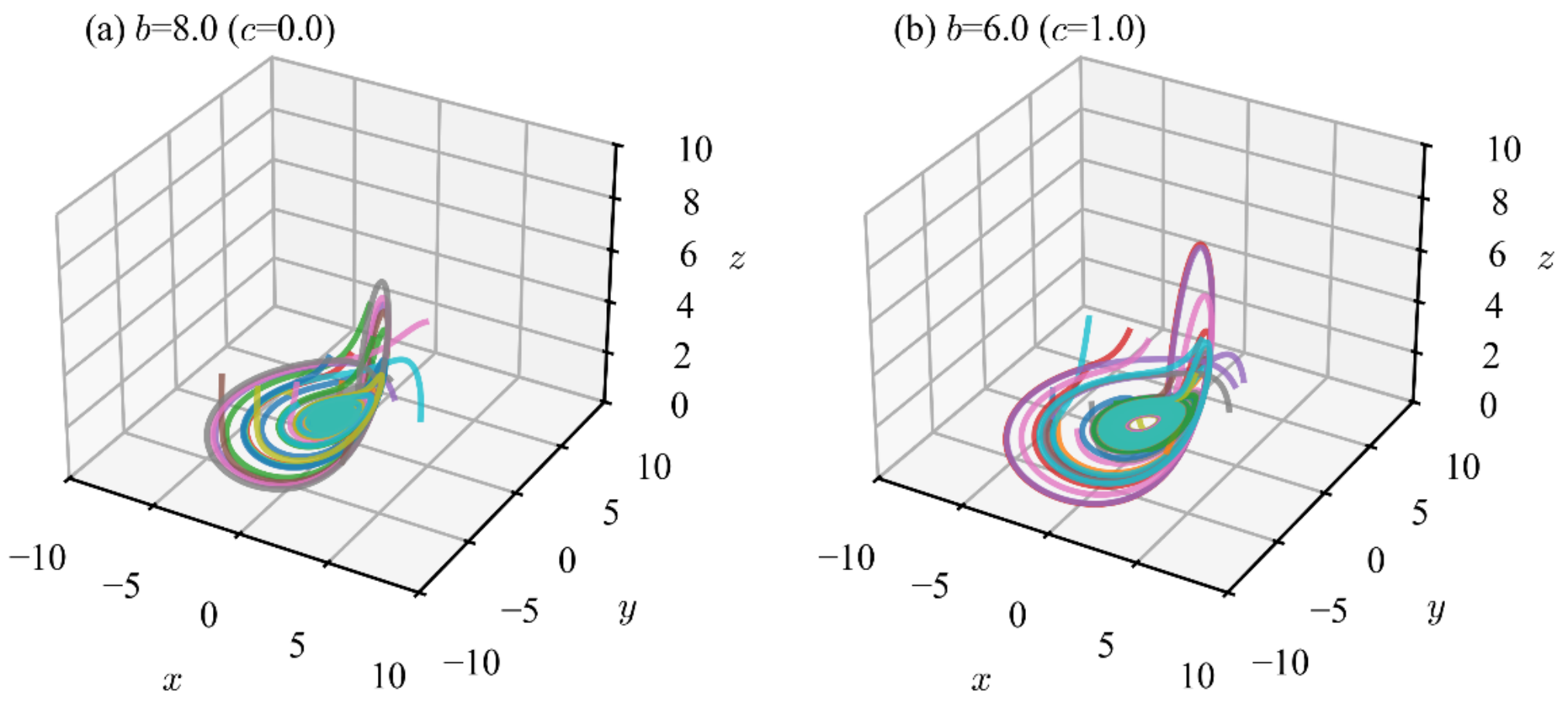

Fig. 15 Trajectories for the training dataset in the Rössler system corresponding to (a) $b = 8$ and (b) $b = 6$.

Using this dataset, we trained ELMC with parameters $N_{in} = 3$, $N_c = 1$, $N_h = 100$, $\alpha = 0.005$, and $\beta = 0.00001$. Fig. 16 shows the time-series predicted by ELMC for $b = 0.2$ ($c = 3.9$), demonstrating that ELMC achieved accurate short-term predictions up to around $t \approx 100$ beyond which the accuracy decreased. Nevertheless, the ELMC successfully reconstructed the attractor, as illustrated in Fig. 17.

A comparison between the bifurcation diagram of the Rössler system predicted by ELMC and that obtained using the fourth-order Runge-Kutta method is presented in Fig. 18. The predicted bifurcation diagram displays several windows of periodic orbits that are absent in the actual diagram; however, the overall bifurcation structure is reproduced well.

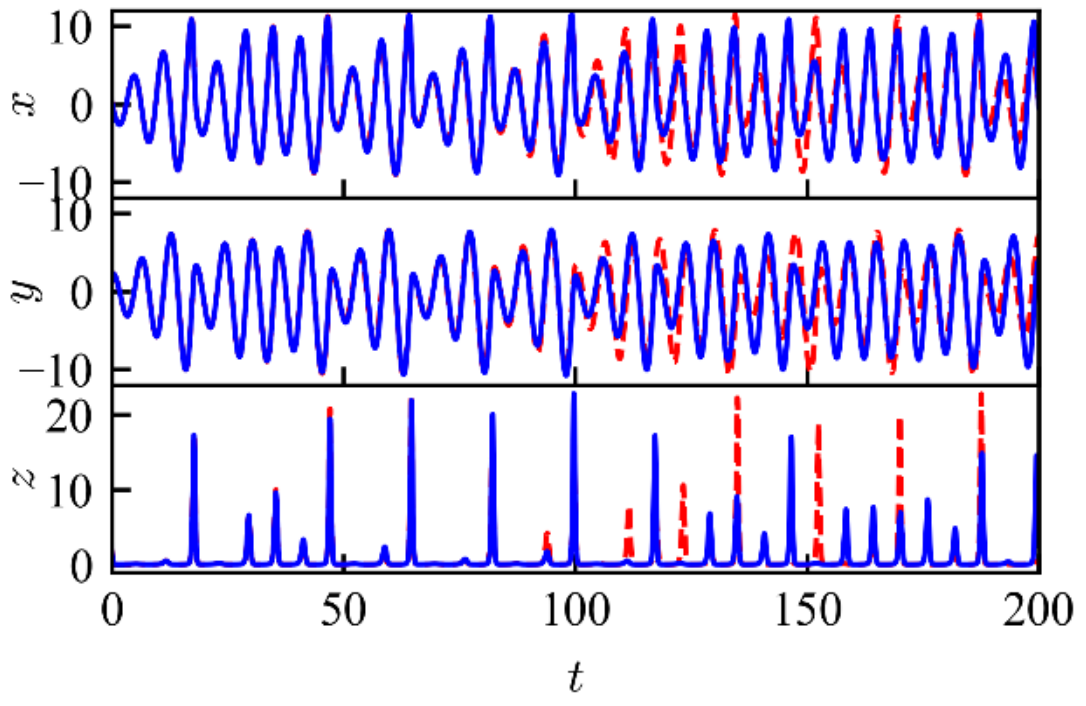

Fig. 16 Comparison of actual (red dotted curve) and predicted (blue solid curve) time-series for the variables of $x$, $y$, and $z$ of the Rössler system at $b = 0.2$, with the same initial values of $(x, y, z)$. The predicted time-series was generated by the trained ELMC with a control input of c = 3.9.

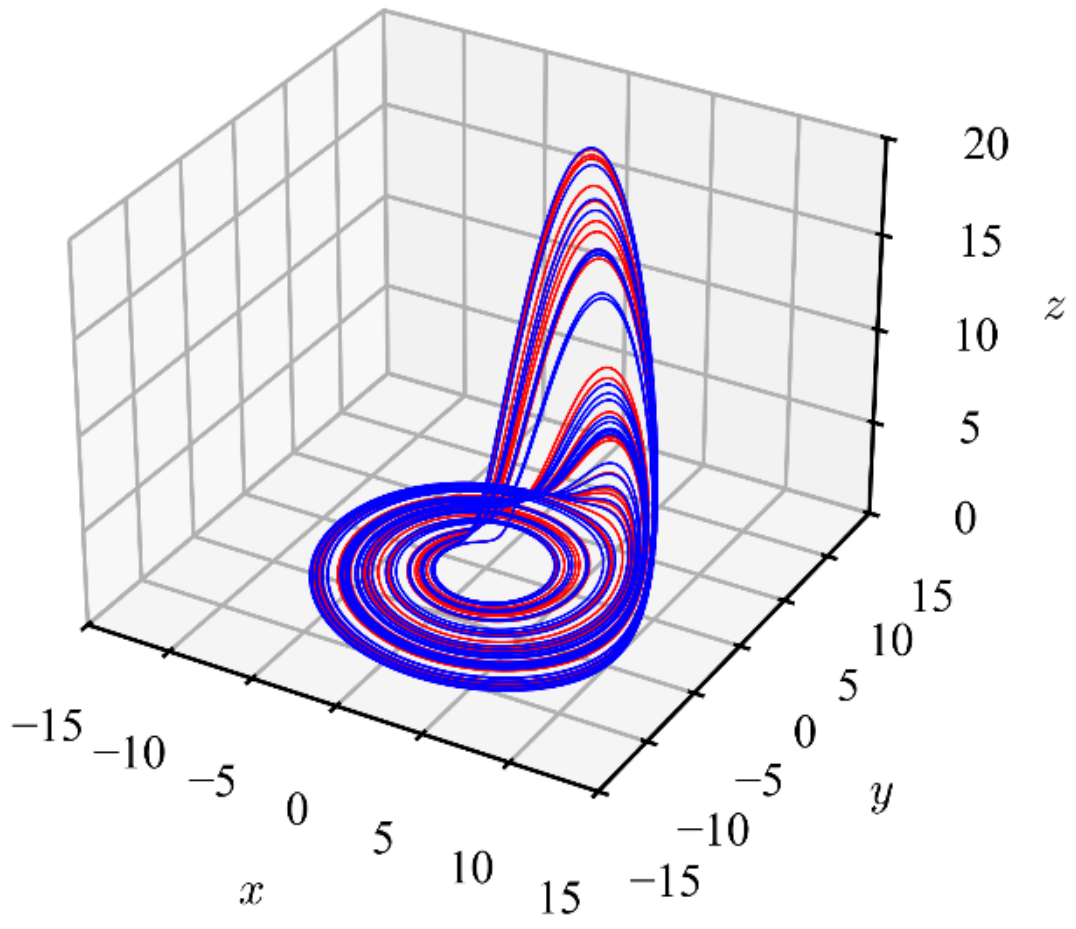

Fig. 17 Rössler attractor obtained by the fourth-order Runge-Kutta method (red curve) and predicted by the ELMC (blue curve) at bifurcation parameter

Given β = 0.00005 and the range of $b$ in the bifurcation diagram in Fig. 18 corresponding to c = 3 to 4, the maximum of $|\delta|$ was estimated as $max(|\delta|) = max(|\beta \boldsymbol{W}^c(c - c_1)|) \sim 0.00005 \times max(|c|) \sim 0.0002$. Therefore, it is reasonable to conclude that the approximations in Eq. (15) also hold for the Rossler system.

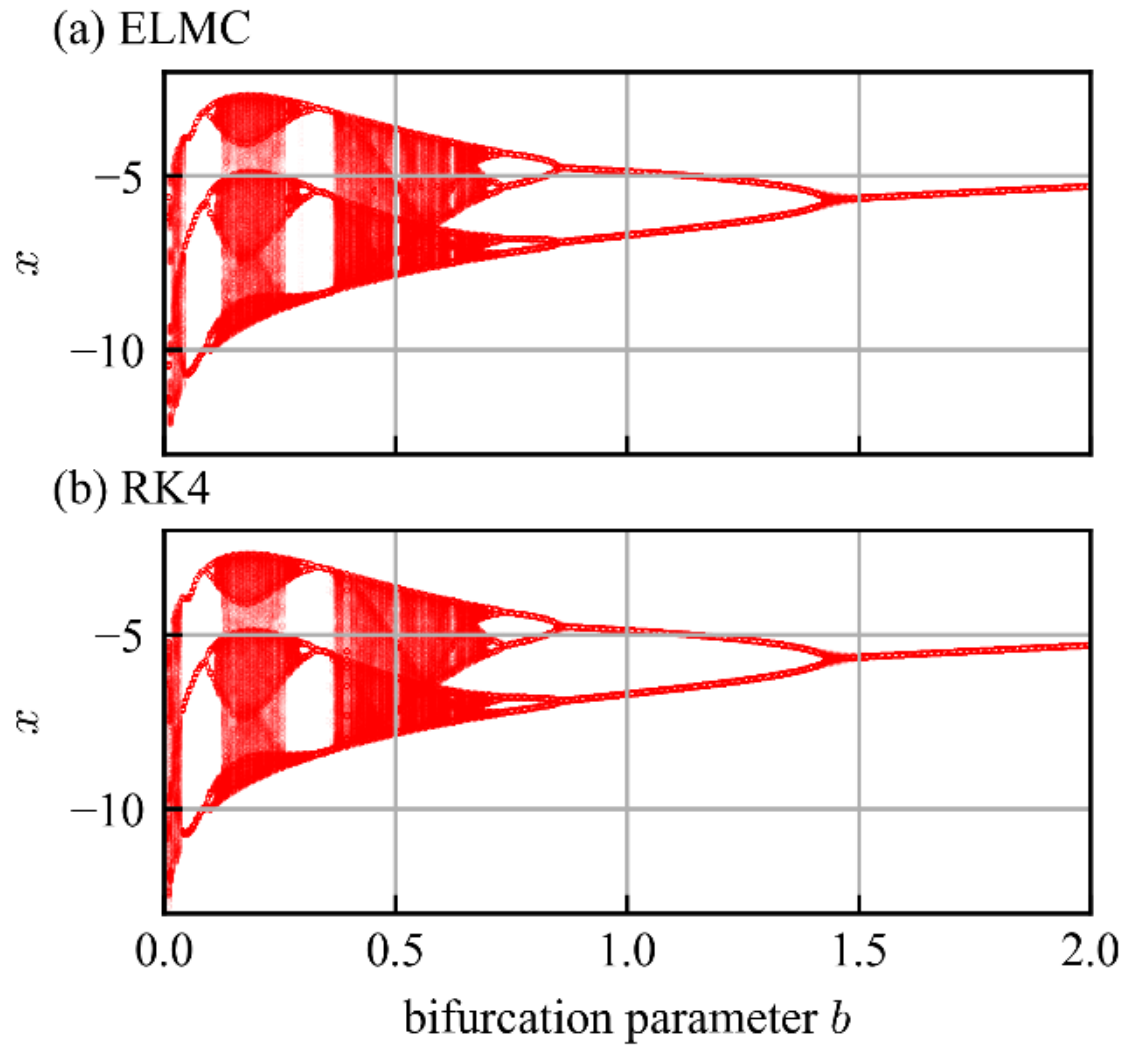

Fig. 18 Bifurcation diagrams of the Rössler system. Variable $x_1$ is plotted On the Poincaré section defined by $y = 0$ and $\dot{y} > 0$. The upper panel indicates the bifurcation diagram predicted by ELMC, whereas the lower one displays the bifurcation diagram obtained by numerical integration using the fourth-order Runge-Kutta method.

## SUMMARY AND DISCUSSIONS

In this study, we extended the conventional extreme learning machine to enhance its predictive capabilities. Supplementary neurons were introduced into the input layer, similar to Kim et al. [11] , for ELM to not only predict the behaviors of dynamical systems but also reproduce a bifurcation structure that extends far beyond the range of learned parameter values. These additional neurons register the control inputs and effectively assimilate the bifurcation parameters of the learned dynamical system.

We applied this augmented model, called ELM with control inputs, to predict dynamic behaviors and reproduce an entire bifurcation structure in both discrete and continuous-time dynamical systems.

By training the ELMC for the time series of the target dynamical systems at only two values of the bifurcation parameter, we demonstrated its performance in replicating dynamical behaviors in regions of the bifurcation parameter that are far from the training domain. Moreover, the ELMC reproduced the global bifurcation structures through appropriate manipulation of the control input.

Concerning the mechanism underlying these remarkable abilities of ELMC, we identified several key factors. First, ELMC learns not the explicit time series of the dynamical system but rather the function governing the dynamical system. In continuous-time dynamical systems, a discretization procedure for numerically solving differential equations, e.g., the fourth order Runge-Kutta, can be considered. Second, if the activation functions of the hidden layer neurons can be approximated well up to the first order of power expansion with respect to the control input, the activation functions remote to the values of the bifurcation parameter used for learning can be reproduced by linear interpolation or extrapolation of the functions used during learning. Moreover, if such a function driving a dynamical system includes a linear bifurcation parameter , a linear functional relationship holds between the bifurcation parameter and control input.

By introducing control inputs into an echo state network [11] , referred to as ESNC, Kim et al. succeeded in predicting dynamical behaviors at parameter values far from the parameter values used for learning and reproduced the global period-doubling bifurcation structures. Then, the question arises: Does the mechanism proposed in the present study adequately explain the predictive capabilities of ESNC studied by Kim et al.? Their ESNC employed continuous-time dynamical systems for the reservoir neurons and utilized discrete-time neurons in the hidden layer. However, the difference is not significant. As demonstrated below, instead of continuous-time neurons, ESNC with discrete-time neurons can yield accurate predictions.

The states of the hidden layer neurons (reservoir neurons) in the ESNC are not solely determined by external inputs but are influenced by past states. The states $\boldsymbol{h}(n)$ are updated according to the following equation:

$$\boldsymbol{h}(n+1) = (1-\gamma)\boldsymbol{h}(n) + \gamma\mathbf{tanh}\left(\boldsymbol{A}\boldsymbol{h}(n) + \alpha\boldsymbol{W}^{\boldsymbol{in}}\boldsymbol{u}(n) + \beta\boldsymbol{W}^{\boldsymbol{c}}\boldsymbol{c}(n) + \boldsymbol{\theta}\right) \tag{32}$$

where $\boldsymbol{A}$ is an adjacency matrix (recurrent connection matrix) and $1-\gamma$ represents a leak. The adjacency matrix $\boldsymbol{A}$ was selected as the connectivity density $\rho = 0.1$, with its elements randomly and uniformly chosen from the interval $[-1, 1]$. Subsequently, $\boldsymbol{A}$ was normalized by dividing it by its largest real-component eigenvalue, followed by multiplication by 0.95. Here on, let us consider the case $\gamma = 1$, then Eq.(32) becomes:

$$\boldsymbol{h}(n+1) = \tanh\left(\boldsymbol{A}\boldsymbol{h}(n) + \alpha\boldsymbol{W}^{\boldsymbol{in}}\boldsymbol{u}(n) + \beta\boldsymbol{W}^{\boldsymbol{c}}\boldsymbol{c}(n) + \boldsymbol{\theta}\right) \tag{33}$$

Given datasets $(\boldsymbol{X}, \boldsymbol{C}, \boldsymbol{D})$, we constructed the state matrix $\boldsymbol{H}$ using (33), and then determined matrix $\boldsymbol{W}^{out}$, which is achieved by minimizing the matrix 2-norm $||\boldsymbol{D} - \boldsymbol{W}^{out}\boldsymbol{H}||^2$. This training procedure is consistent with that of ELMC. Nevertheless, unlike the case of ELMC, $\boldsymbol{h}(n+1)$ depends on $\boldsymbol{h}(n)$ in ESNC. To determine $\boldsymbol{W}^{out}$, we multiply both sides of Eq. (33) by $W^{out}$, thereby obtaining the following autonomous system:

$$\begin{aligned} \boldsymbol{x}(n+1) &= \boldsymbol{W}^{out}\tanh\left(\boldsymbol{A}\boldsymbol{h}(n) + \alpha\boldsymbol{W}^{in}\boldsymbol{u}(n) + \beta\boldsymbol{W}^{c}\boldsymbol{c} + \boldsymbol{\theta}\right) \\ \boldsymbol{h}(n+1) &= \mathbf{tanh}\left(\boldsymbol{A}\boldsymbol{h}(n) + \alpha\boldsymbol{W}^{in}\boldsymbol{u}(n) + \beta\boldsymbol{W}^{c}\boldsymbol{c} + \boldsymbol{\theta}\right) \end{aligned} \tag{34}$$

Here, the temporal index $n$ of the control input $\boldsymbol{c}$ is omitted as in the case of ELMC.

Let us apply ESNC to the Lorenz system. Using the parameters $N_h = 100$, $\alpha = 0.005$, $\beta = 0.0007$ and employing the same dataset used in Section VI.A, we proceed to train the ESNC. To execute the ESNC post-learning process, not only the initial state variable $\boldsymbol{x}(0)$ but also the initial internal state variable $\boldsymbol{h}(0)$ of the reservoir neurons is required. To this end, we used the values of the internal state variables at the end of the learning phase. Fig. 19 illustrates the trajectory at $r = 28$

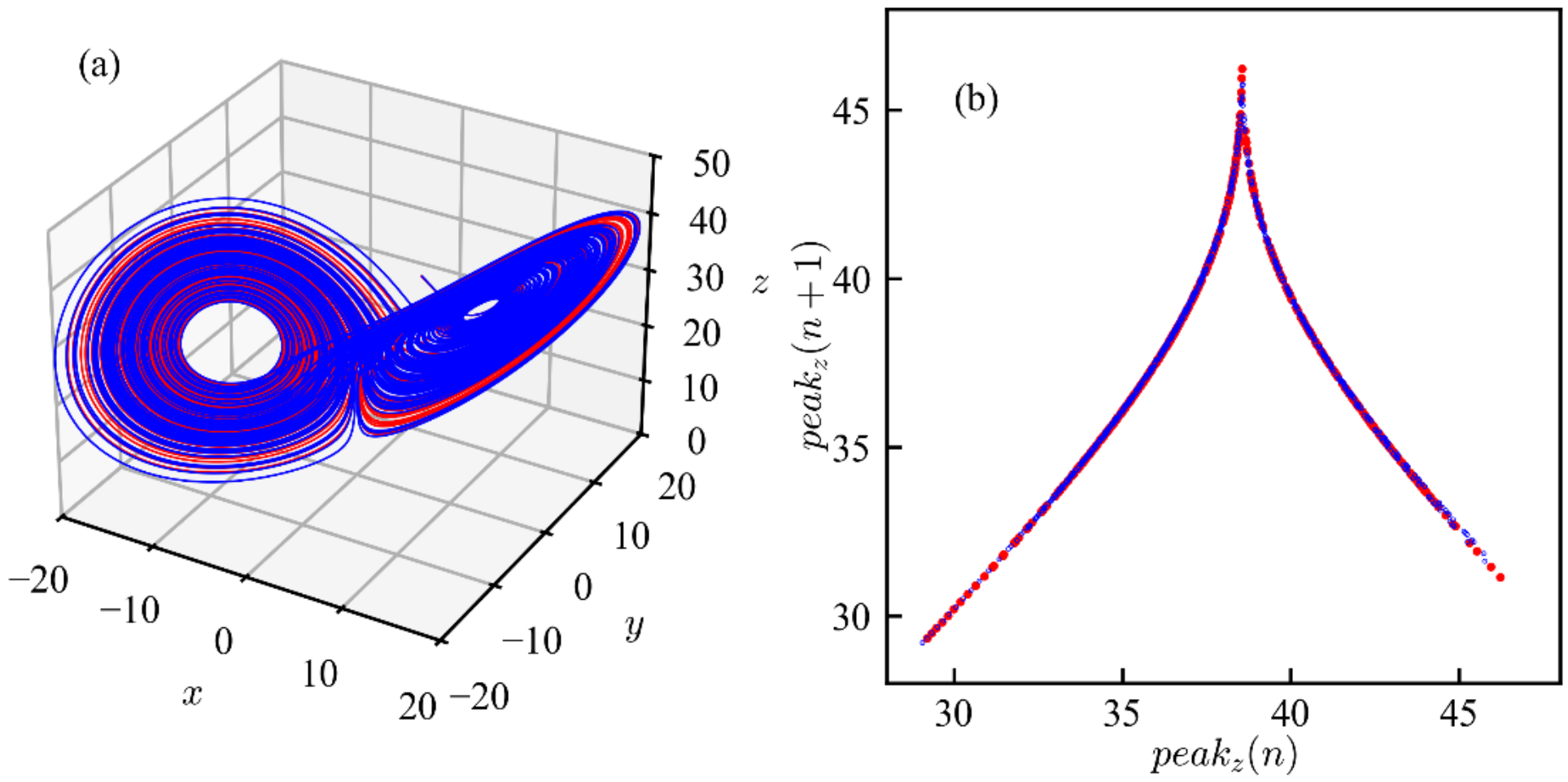

Fig. 19 (a) Comparison of a strange attractor trajectory at $r = 28$. The red curve denotes the trajectory obtained using the fourth-order Runge-Kutta method (RK4), whereas the blue curve depicts the trajectory predicted by the echo state network with a control input (ESNC) of $c = 5$. (b) Lorenz plots for the trajectories shown in (a). The red and blue dots correspond to the RK4 and ESNC trajectories, respectively.

predicted by the trained ESNC and its Lorenz plot. The results demonstrate that the ESNC with discrete-time reservoir neurons, effectively reproduces the dynamical structure of the Lorenz system at $r = 28$. The bifurcation diagram predicted by ESNC is also reproduced, as shown in Fig. 20.

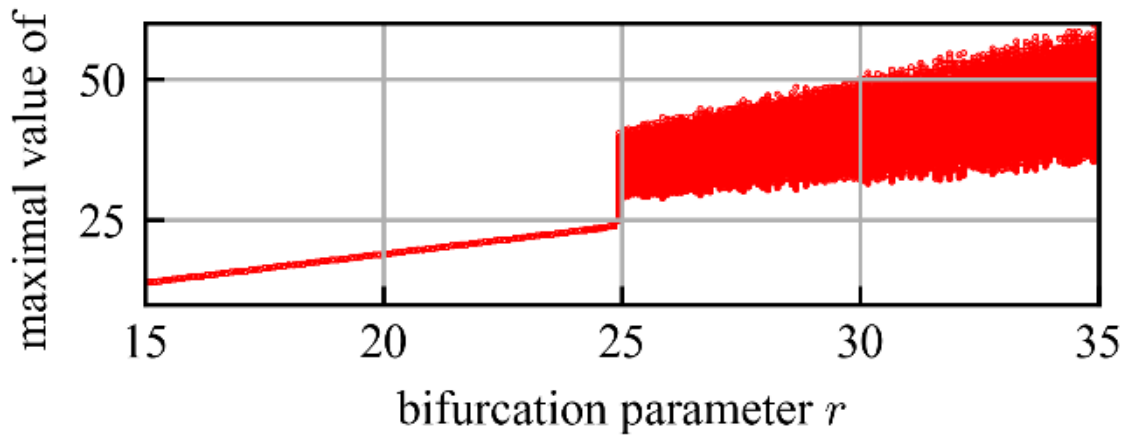

Fig. 20 Bifurcation diagram inferred by ESNC. The local maxima of $z$ of the trajectories predicted by ESNC are plotted as a function of the bifurcation parameter $r$.

Now, let us examine whether the proposed mechanism can elucidate the predictive capabilities of ESNC. Eq. (34) suggests the possibility of defining a function $f^{ESNC}(c, \boldsymbol{u})$, that approximates the function governing the target dynamical system, which can construct a dynamical system $\boldsymbol{u}(n+1) = f^{ESNC}\left(c, \boldsymbol{u}(n)\right)$. However, because of the dependence of the state $\boldsymbol{h}(n+1)$ in Eq. (34) on both $\boldsymbol{u}(n)$ and the preceding state $\boldsymbol{h}(n)$, it is impossible to define such a function based solely on $\boldsymbol{x}(n)$ and $c$. If the time dependence of the recurrent term in Eq. (34) and the dependency on the initial values $\boldsymbol{h}(0)$ are negligible, defining $f^{ESNC}(c, \boldsymbol{u})$ becomes feasible. This issue will be addressed in future studies.

Acknowledgments

I.T. was partially supported by the JST Strategic Basic Research Programs (Symbiotic Interaction: Creation and Development of Core Technologies Interfacing Human and Information Environments, CREST Grant Number JPMJCR17A4.

Declarations:

Conflict of interest : No conflict of interest

Author contributions

Conceptualization, S.T, A.Y., T.N. and I.T.; formulation and computer simulation, S.T. and A.Y.; writing—original draft preparation, S. T. and A.Y.; writing—review and editing, I.T. and T.N. All authors have read and agreed to the published version of the manuscript.

Data Availability Statements

The data that support the findings of this study are available from the corresponding author upon reasonable request.